%% file: HIG-16-016_temp.tex
\pdfoutput=1

\documentclass[11pt,twoside,a4paper,cmspaper,final,collab]{cms-tdr}
\def\svnVersion{391163}\def\cmsCernNoTag{CERN-EP/2016-240}\def\cmsCernDate{\today}\def\cmsMessage{Published in the Journal of High Energy Physics as \href{http://dx.doi.org/10.1007/JHEP02(2017)135}{\doi{10.1007/JHEP02(2017)135.}}}

\begin{document}\cmsNoteHeader{HIG-16-016}

\hyphenation{had-ron-i-za-tion}
\hyphenation{cal-or-i-me-ter}
\hyphenation{de-vices}
\RCS$Revision: 391148 $
\RCS$HeadURL: svn+ssh://svn.cern.ch/reps/tdr2/papers/HIG-16-016/trunk/HIG-16-016.tex $
\RCS$Id: HIG-16-016.tex 391148 2017-03-03 15:25:59Z nckw $
\newlength\cmsFigWidth
\setlength\cmsFigWidth{0.4\textwidth}
\providecommand{\cmsTable}[1]{\resizebox{\textwidth}{!}{#1}}

\newcommand{\zllh}{\ensuremath{\Z(\ell\ell)\PH}\xspace}
\newcommand{\epem}{\ensuremath{\Pep\Pem}\xspace}
\newcommand{\ptll}{\ensuremath{\pt^{\ell\ell}}\xspace}
\newcommand{\mpmm}{\ensuremath{\PGmp\PGmm}\xspace}
\newcommand{\Zmmjets}{\ensuremath{\Z(\mu^{+}\mu^{-})\text{+jets}}}
\newcommand{\Zeejets}{\ensuremath{\Z(\Pep\Pem)\text{+jets}}}
\newcommand{\DYjets}{\ensuremath{\Z/\gamma^{*}(\ell^{+}\ell^{-})\text{+jets}}}
\newcommand{\Zvvjets}{\ensuremath{\Z(\nu\nu)  \text{+jets}}}
\newcommand{\Wlvjets}{\ensuremath{\PW(\ell\nu)          \text{+jets}}}
\newcommand{\Wevjets}{\ensuremath{\PW(\Pe\nu)          \text{+jets}}}
\newcommand{\Wmvjets}{\ensuremath{\PW(\mu\nu)          \text{+jets}}}
\newcommand{\Wtvjets}{\ensuremath{\PW(\tau\nu)          \text{+jets}}}
\newcommand{\zjets}{\ensuremath{\Z\text{+jets}}}
\newcommand{\wjets}{\ensuremath{\PW\text{+jets}}}
\newcommand{\vjets}{\ensuremath{\mathrm{V}\text{+jets}}}
\newcommand{\phojets}{\ensuremath{\gamma\text{+jets}}}
\newcommand{\higgsbrobs}{\ensuremath{0.24}}
\newcommand{\higgsbrobsninety}{\ensuremath{0.20}}
\newcommand{\higgsbrexp}{\ensuremath{0.23}}
\newcommand{\mjj}{\ensuremath{m_{\mathrm{jj}}}}
\newcommand{\brinv}{\ensuremath{\mathcal{B}(\PH\to \text{inv})}\xspace}
\newcommand{\Ginv}{\ensuremath{{\Gamma}_{\text{inv}}}}
\newcommand{\Gsm}{\ensuremath{{\Gamma}_{\mathrm{SM}}}}
\newcommand{\sigbrinv}{\ensuremath{\sigma\,\brinv/\sigma(\mathrm{SM})}\xspace}
\newcommand{\sighatbrinv}{\ensuremath{\sigma\,\hat{\mathcal{B}}(\PH\to \text{inv})/\sigma(\mathrm{SM})}}
\newcommand{\detajj}{\ensuremath{{\Delta\eta(\mathrm{j_{1},j_{2}})}}}
\newcommand{\ptjj}{\ensuremath{\pt^{\mathrm{j_{1}}},~\pt^{\mathrm{j_{2}}}}}
\newcommand{\mindphi}{\ensuremath{{\min\Delta\phi(\ptvecmiss,\mathrm{j})}}}
\newcommand{\dphijm}{\ensuremath{{\Delta\phi(\ptvecmiss,\mathrm{j})}}}
\newcommand{\mll}{\ensuremath{m_{\mathrm{\ell\ell}}}}
\providecommand{\mt}{\ensuremath{m_{\mathrm{T}}}}
\newcommand{\zzllvv}{\ensuremath{\Z\Z \to \ell\ell\nu\nu}}
\newcommand{\wzlvll}{\ensuremath{\mathrm{WZ} \to \ell\nu \ell\ell}}
\newcommand{\dyee}{\ensuremath{\Z/\gamma^*\to \mathrm{\Pe^+\Pe^-}}}
\newcommand{\dymm}{\ensuremath{\Z/\gamma^*\to\mu^+\mu^-}}
\newcommand{\dyll}{\ensuremath{\Z/\gamma^*\to \ell^+\ell^-}}
\newcommand{\zee}{\ensuremath{\Z\to\mathrm{e^+e^-}}}
\newcommand{\vjj}{\ensuremath{\textrm{V}(\mathrm{jj})}}
\newcommand{\zbb}{\ensuremath{\Z(\mathrm{b\bar{b}})}}
\newcommand{\zvv}{\ensuremath{\Z(\nu\nu)}}
\newcommand{\zmm}{\ensuremath{\Z\to\mu^+\mu^-}}
\newcommand{\zll}{\ensuremath{\Z(\ell^+\ell^-)}}
\newcommand{\wenu}{\ensuremath{\PW\to \Pe\nu}}
\newcommand{\wmunu}{\ensuremath{\PW\to \mu\nu}}
\newcommand{\wtaunu}{\ensuremath{\PW\to \tau\nu}}
\providecommand{\NA}{\ensuremath{\text{---}\xspace}}

\cmsNoteHeader{HIG-16-016}
\title{Searches for invisible decays of the Higgs boson in pp collisions at $\sqrt{s}=7,$ 8, and 13\TeV}

\date{\today}

\abstract{
Searches for invisible decays of the Higgs boson are presented.
The data collected with the CMS detector at the LHC correspond to integrated luminosities of 5.1, 19.7, and 2.3\fbinv at centre-of-mass
energies of 7, 8, and 13\TeV, respectively.  The search channels target Higgs boson production via
gluon fusion, vector boson fusion, and in association with a vector boson. Upper limits are placed on the branching
fraction of the Higgs boson decay to invisible particles, as a function of the assumed production cross sections. The combination of all channels, assuming
standard model production, yields an observed (expected) upper limit on the invisible branching fraction of
\higgsbrobs~(\higgsbrexp) at the 95\% confidence level. The results are also interpreted in the context of Higgs-portal dark matter models.
}

\hypersetup{%
pdfauthor={CMS Collaboration},%
pdftitle={Searches for invisible decays of the Higgs boson in pp collisions at sqrt(s) = 7, 8, and 13 TeV},%
pdfsubject={CMS},%
pdfkeywords={CMS, physics, Higgs, invisible decays}}

\maketitle

\section{Introduction}

The Higgs boson (H) discovery and the study of its properties
by the ATLAS and CMS Collaborations~\cite{Aad:2012tfa,CMSPaperCombination,Chatrchyan:2013lba} at
the CERN LHC have placed major constraints on potential models of new physics beyond
the standard model (SM). Precision measurements of the couplings of the
Higgs boson from a combination of the 7 and 8\TeV ATLAS and CMS data sets indicate a
very good agreement between the measured properties of the Higgs boson and the SM predictions~\cite{CMS-PAS-HIG-15-002}.
In particular, these
measurements provide indirect constraints on additional contributions to the Higgs boson
width from non-SM decay processes. The resulting indirect upper limit on the Higgs boson
branching fraction to non-SM decays is 0.34 at the 95\% confidence level (CL)~\cite{CMS-PAS-HIG-15-002}.

A number of models for physics beyond the SM allow for invisible decay modes of the
Higgs boson, such as decays to neutralinos in supersymmetric models~\cite{Belanger:2001am} or
graviscalars in models with extra spatial dimensions~\cite{Giudice:2000av,Dominici:2009pq}. More generally, invisible Higgs boson decays
can be realised through interactions between the Higgs boson and dark matter (DM)~\cite{SHROCK1982250}.
In Higgs-portal models~\cite{Baek:2012se,Djouadi:2011aa,Djouadi:2012zc,Beniwal:2015sdl},
the Higgs boson acts as a mediator between SM and
DM particles allowing for direct production of DM at the LHC. Furthermore, cosmological
models proposing that the Higgs boson played a central role in the evolution of the
early universe motivate the study of the relationship between the Higgs boson and DM~\cite{Servant:2013uwa,Cohen:2012zza}.

Direct searches for invisible decays of the Higgs boson increase the sensitivity to the
invisible Higgs boson width beyond the indirect constraints. The typical signature at the LHC is a large missing
transverse momentum recoiling against a distinctive visible system. Previous searches by the
ATLAS and CMS Collaborations have targeted Higgs boson production in association with a vector boson
(VH, where V denotes W or Z)~\cite{Aad:2014iia,Aad:2015uga,Chatrchyan:2014tja,Chatrchyan:2014tja} or with jets consistent with a
vector boson fusion (VBF, via $\PQq\PQq\to\PQq\PQq\PH$) topology~\cite{Aad:2015txa,Chatrchyan:2014tja}.
A combination of direct searches for invisible Higgs boson decays in qqH and VH production, by the ATLAS Collaboration,
yields an upper limit of 0.25 on the Higgs boson invisible branching fraction, \brinv, at the 95\% confidence level~\cite{Aad:2015pla}.
Additionally, searches by the ATLAS Collaboration for DM in events with missing transverse momentum accompanied by jets have been interpreted in the context of Higgs boson production via
gluon fusion and subsequent decay to invisible particles~\cite{Aad:2015zva}.

\begin{figure}[hbt]
\centering
\includegraphics[width=0.32\textwidth]{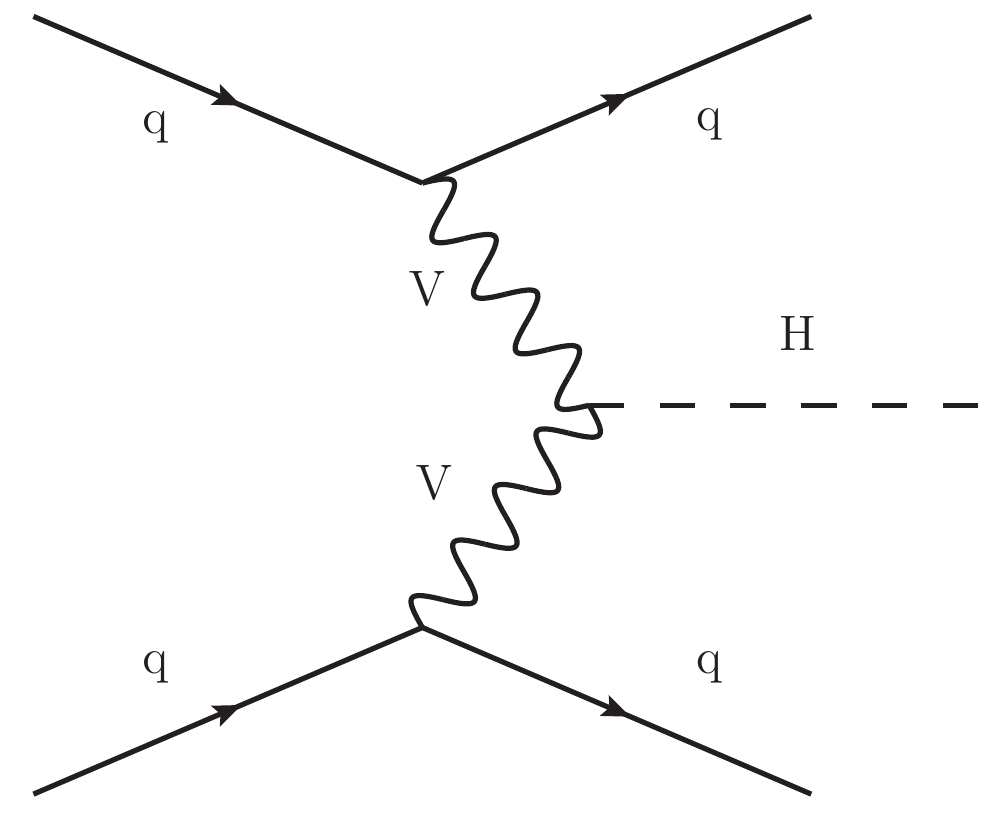}
\includegraphics[width=0.4\textwidth]{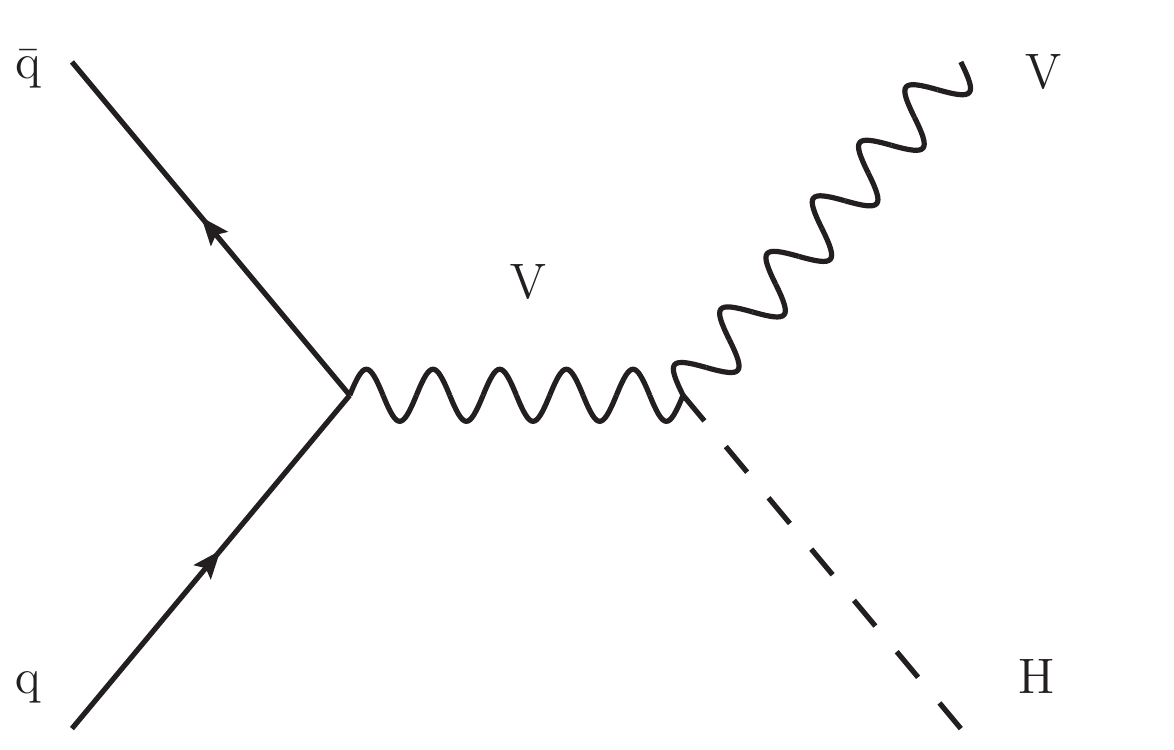}\\
\includegraphics[width=0.4\textwidth]{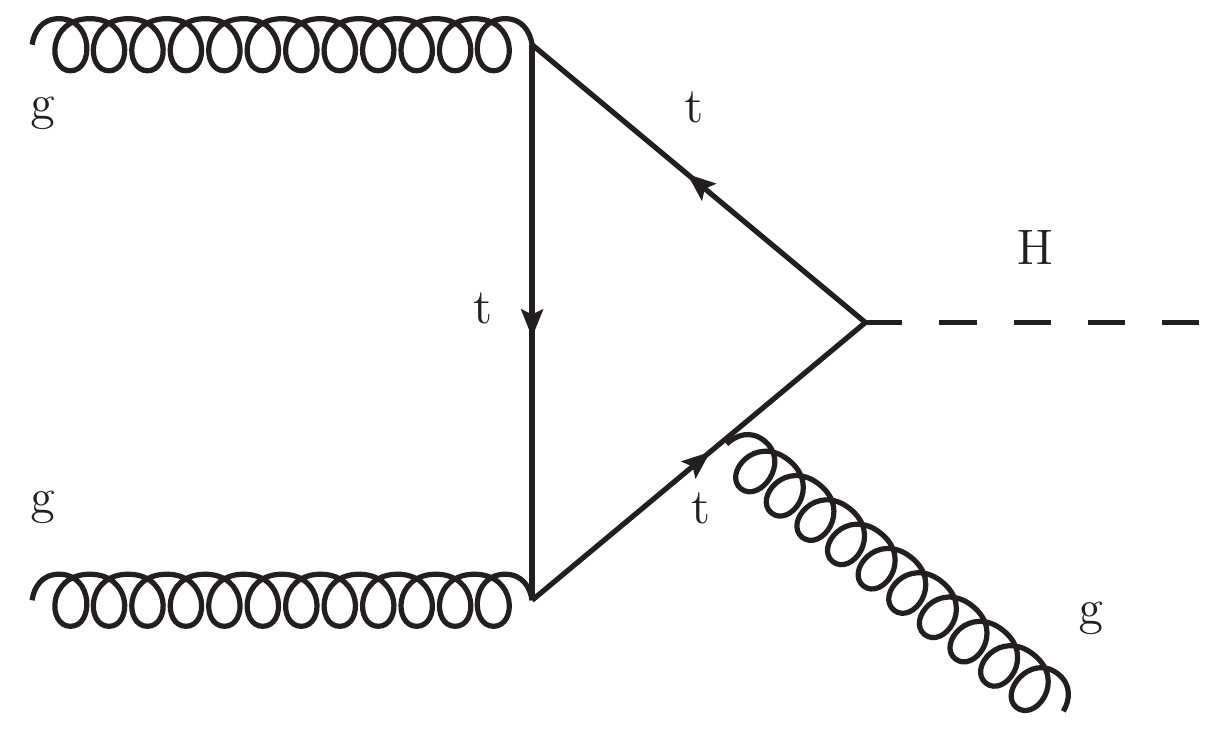}
  \caption{Feynman diagrams for the three production processes targeted in the search for invisible Higgs boson decays:
 (upper left) $\PQq\PQq\to\PQq\PQq\PH$, (upper right) $\qqbar\to\mathrm{V}\PH$, and (bottom) $\Pg\Pg\to\Pg\PH$. \label{fig:production_diagrams}}
\end{figure}

\begin{figure}[hbt]
\centering
  \includegraphics[width=0.49\textwidth]{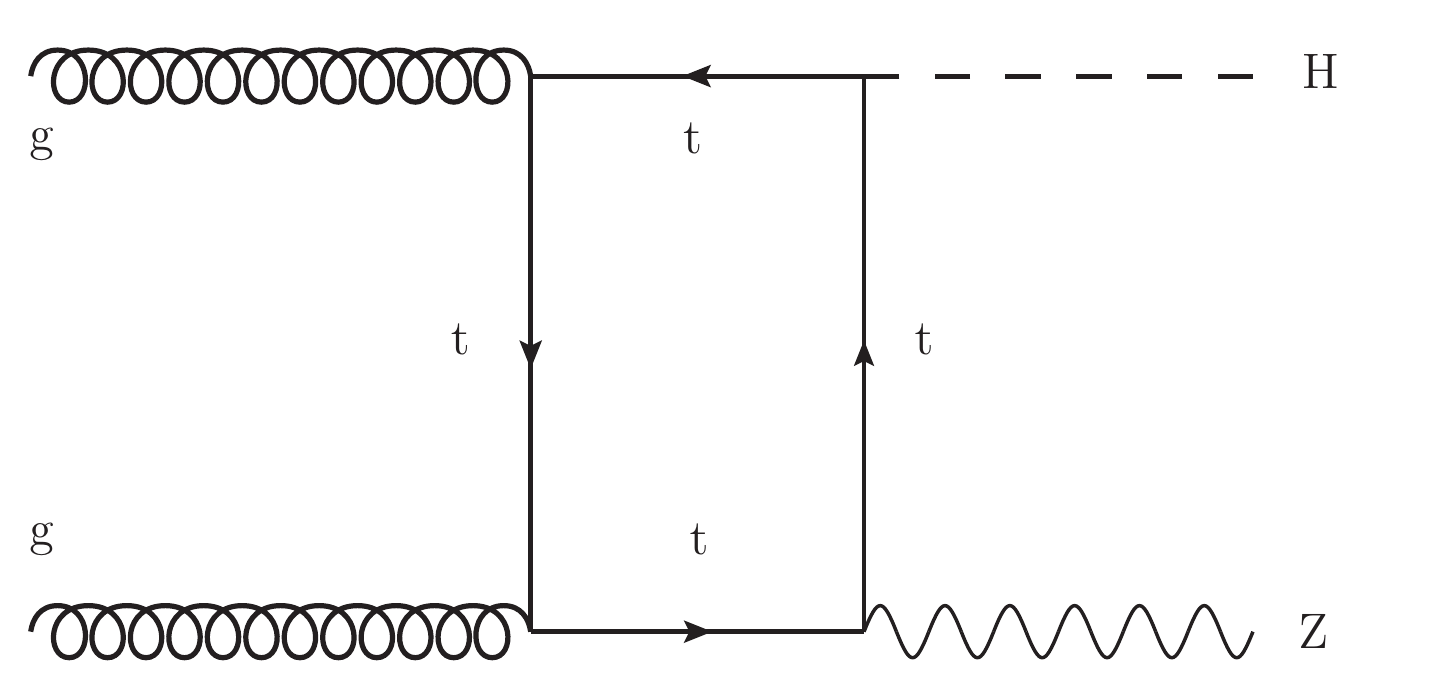} \includegraphics[width=0.49\textwidth]{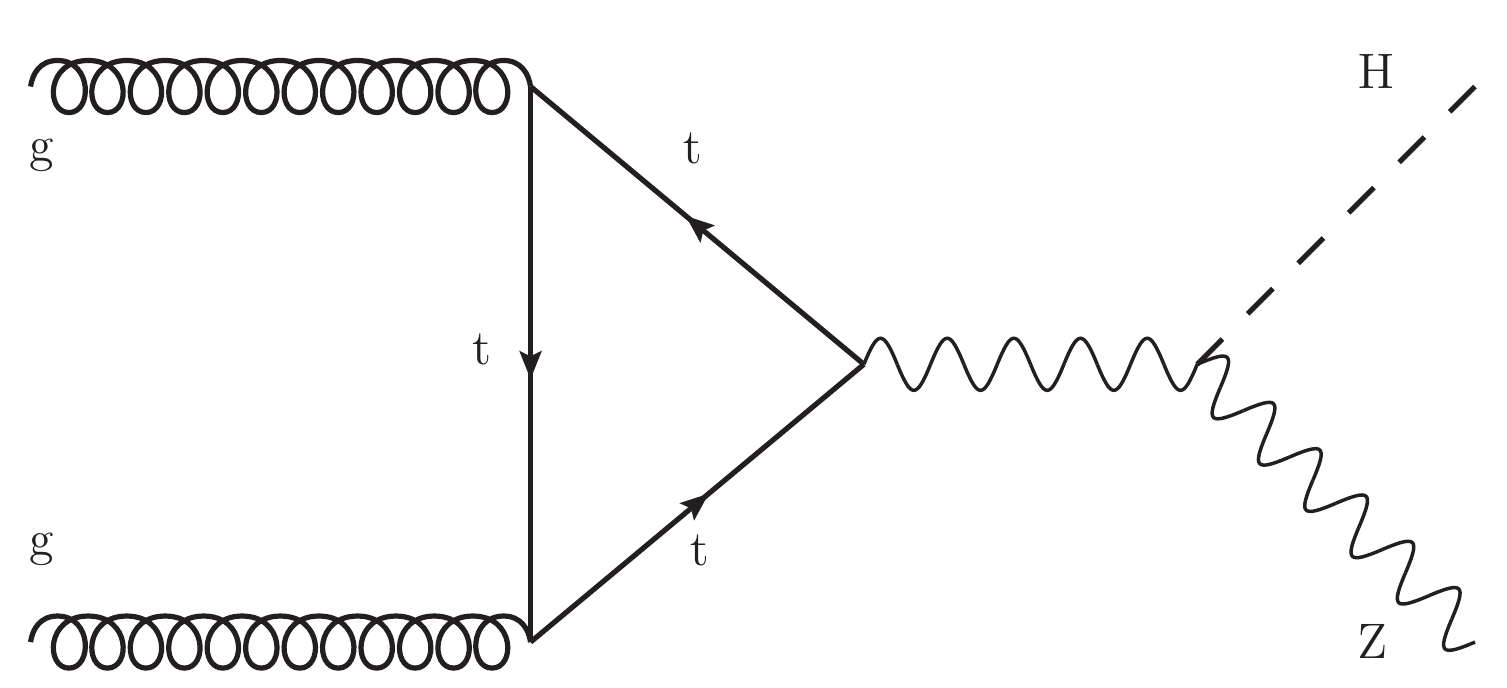} \caption{Feynman diagrams for the gg$\to$ZH production processes involving a coupling between (left) the top quark
  and the Higgs boson or (right) the Z and Higgs bosons.\label{fig:production_ggZH}}
\end{figure}

In this paper, results from a combination of searches for invisible decays of the
Higgs boson using data collected during 2011, 2012, and 2015 are presented.
The searches target the qqH, VH, and ggH production modes. The searches for
the VH production mode include searches targeting ZH production, in which the Z boson decays to a pair of leptons (either $\epem$ or
$\mpmm$) or $\bbbar$, and searches for both the ZH and WH production modes, in which the
W or Z boson decays to light-flavour jets.
Additional sensitivity is achieved in this analysis by including a search targeting gluon fusion production where the Higgs boson
is produced accompanied by a gluon jet ($\Pg\Pg\to\Pg\PH$). The diagrams for the qqH, VH, and ggH Higgs boson production processes are shown
in Fig.~\ref{fig:production_diagrams}. The contribution to ZH production from gluon fusion ($\Pg\Pg\to\Z\PH$), as shown in
Fig.~\ref{fig:production_ggZH}, is included in this analysis.
When combining the searches to determine an upper limit on \brinv SM production cross sections are assumed,
consistent with the measured Higgs boson production rates~\cite{CMS-PAS-HIG-15-002}. In addition,
upper limits on \brinv assuming non-SM production cross sections are provided.

This paper is structured as follows: a brief overview of the CMS detector
and event reconstruction is given in Section~\ref{sec:cms}, and the data sets and simulation used for
the searches are presented in Section 3.
In Section 4, the strategy for each search included in the combination is described,
and in Section 5 the results of the searches are presented and interpreted in terms of
upper limits on \brinv and DM-nucleon interaction cross sections. Finally, a summary is presented in Section 6.

\section{The CMS detector and object reconstruction\label{sec:cms}}

The CMS detector is a
multipurpose apparatus optimised to study high transverse momentum (\pt) physics processes in
proton-proton and heavy-ion collisions.  A superconducting solenoid occupies its
central region, providing a magnetic field of 3.8\unit{T} parallel to the beam
direction. Charged-particle trajectories are measured by the silicon pixel and
strip trackers, which cover a pseudorapidity region of $\abs{\eta} < 2.5$. A
lead tungstate crystal electromagnetic calorimeter (ECAL), and a
brass and scintillator hadron calorimeter (HCAL) surround the tracking volume and
cover $\abs{\eta} < 3$. The steel and quartz-fibre Cherenkov hadron forward
calorimeter extends the coverage to $\abs{\eta} < 5$.  The muon system consists
of gas-ionisation detectors embedded in the steel flux-return yoke outside the
solenoid, and covers $\abs{\eta} < 2.4$. The first level of the CMS trigger
system, composed of custom hardware processors, is designed to select the most
interesting events in less than 4\mus, using information from the calorimeters
and muon detectors. The high-level trigger processor farm then further reduces
the event rate to less than 1\unit{kHz}. A more detailed description of the CMS detector,
together with a definition of the coordinate system used and the relevant kinematic variables,
can be found in Ref.~\cite{Chatrchyan:2008zzk}.

Objects are reconstructed using the CMS particle-flow (PF)
algorithm~\cite{CMS-PAS-PFT-09-001,CMS-PAS-PFT-10-001}, which optimally combines information from the
various detector components to reconstruct and identify individual particles.
The interaction vertex with the maximum value of $\sum_{i} (\pt^{i})^{2}$, where $\pt^{i}$ is the
transverse momentum of the $i$th track associated with the vertex, is selected as the primary
vertex for the reconstruction of these objects.

Jets are reconstructed by clustering PF candidates,
using the anti-\kt algorithm~\cite{Cacciari:2008gp} with a distance parameter of 0.5\,(0.4) for the 7 and 8\,(13)\TeV data set.
Analyses exploring Lorentz-boosted hadronic objects employ large-radius jets, clustered using the
Cambridge--Aachen algorithm~\cite{cajets} at 8\TeV and the
anti-\kt algorithm at 13\TeV, each with a distance parameter of 0.8.
The combined secondary vertex algorithm is used to identify
jets originating from b quarks (b jets)~\cite{Chatrchyan:2012jua,btag8TeVPAS,CMS-PAS-BTV-15-001}. The selection used is roughly 70\% efficient for
b jets with $\pt>30$\GeV.

The jet momentum is corrected to account for contamination from additional interactions in the same bunch crossing (pileup, PU)
based on the event energy density scaled proportionally to the jet area~\cite{Cacciari:2011ma}.
Calibrations based on simulation and control samples in data are applied to correct the absolute scale of
the jet energy~\cite{Khachatryan:2016kdb}. The jets are further subjected to a standard set of identification
criteria~\cite{jec}. All jets are required to have $\pt>30$\GeV and
$\abs{\eta}<4.7$, unless stated otherwise.

The missing transverse momentum vector \ptvecmiss is defined as the projection on the
plane perpendicular to the beams of the negative vector sum of the momenta of
all PF candidates in the event. The magnitude of \ptvecmiss is referred to as \ETm.
Dedicated quality filters are applied for tracks, muons, and other physics objects to remove events
with large misreconstructed \ETm.

Electron ($\Pe$), photon ($\gamma$), and muon ($\mu$) candidates are required to be within the relevant
detector acceptances of $\abs{\eta}< 2.5$ ($\Pe/\gamma$) and $\abs{\eta}<2.4$ ($\mu$). Electron and photon candidates in the
transition region between the ECAL barrel and endcap ($1.44 < \abs{\eta} < 1.57$) are not considered because the reconstruction of electrons and
photons in this region is not optimal.
Details of the electron, photon, and muon reconstruction algorithms and their performance
can be found in Refs.~\cite{Khachatryan:2015hwa}, \cite{Khachatryan:2015iwa}, and~\cite{Chatrchyan:2012xi}, respectively.

Lepton isolation is based on the sum of the \pt of additional PF candidates
in a cone of radius $R=\sqrt{\smash[b]{(\Delta\eta)^2+(\Delta\phi)^2}} = 0.4$
around each lepton,
where $\Delta\phi$ and $\Delta\eta$ are the
differences in azimuthal angle (in radians) and pseudorapidity between the
lepton and each particle in the sum, respectively.
The isolation sum is required to be smaller than 15\% (12\%) of the electron (muon) \pt.
In order to reduce the dependence of the isolation variable on the
number of PU interactions, charged hadrons are included in the sum only
if they are consistent with originating from the selected primary vertex of the event.
To further correct for the contribution of neutral particles from PU events to
the isolation sum in the case of electrons, the median transverse energy density, determined on an
event-by-event basis as described in Ref.~\cite{FASTJET}, is subtracted from
the sum. For muons the correction is made by subtracting half the sum of the transverse momenta of
charged particles that are inside the cone and not associated with the primary vertex. The factor of one half accounts for
the expectation that there are half the number of neutral particles
as charged particles within the cone.

Details of the reconstruction of $\tau$ leptons can be found in Ref.~\cite{Khachatryan:2015dfa}. The sum of the transverse momenta
of all PF candidates within a cone of radius $\Delta R<0.3$ around the $\tau$ candidates is required to be less than 5\GeV.

For the purposes of event vetoes, a set of electron, photon, muon, and $\tau$-lepton identification and isolation criteria are applied as
defined by the ``loose'' selections in Refs.~\cite{Khachatryan:2015hwa},~\cite{Khachatryan:2015iwa},~\cite{Chatrchyan:2013sba},
and~\cite{Khachatryan:2015dfa}, respectively. To veto an event the electron, photon, or muon must have $\pt>10$\GeV and fall
within the detector acceptance described above, while a $\tau$-lepton must have $\pt>15$\GeV and $\abs{\eta}<2.3$.
These vetoes suppress backgrounds from leptonic decays of electroweak (EW) backgrounds and allow orthogonal control regions.

\section{Data samples and simulation}

The data used for the analyses described here comprise pp collisions collected with the CMS detector in
the 2011, 2012, and 2015 data-taking periods of the LHC. The integrated
luminosities are 4.9, 19.7, and 2.3\fbinv at centre
of mass energies of 7, 8, and 13\TeV, respectively. The uncertainties
in the integrated luminosity measurements are 2.2\%, 2.6\%, and 2.7\% at 7~\cite{lumi7TeVPAS}, 8~\cite{lumi8TeVPAS},
and 13\TeV~\cite{lumi13TeVPAS}, respectively.

Simulated ggH and qqH events are generated with {\POWHEG1.0} ({\POWHEG2.0})~\cite{Alioli:2010xd,Nason:2009ai,Alioli:2008tz}
interfaced with {\PYTHIA6.4}~\cite{Sjostrand:2006za} ({\PYTHIA8.1}~\cite{Sjostrand:2007gs}) at
7 and 8\,(13)\TeV.
The inclusive cross section for ggH production is calculated to next-to-next-to-next-to-leading
order (N$^3$LO) precision in quantum chromodymanics (QCD) and next-to-leading order (NLO) in EW
theory~\cite{Anastasiou:2014vaa}. The qqH inclusive cross section calculation uses next-to-next-to-leading order (NNLO) QCD
and NLO EW precision~\cite{Heinemeyer:2013tqa}.
In the 8\TeV sample, the \pt distribution of the Higgs boson in the
ggH process is reweighted to match the NNLO plus
next-to-next-to-leading-logarithmic (NNLL) prediction from
\textsc{HRes2.1} ~\cite{deFlorian:2012mx,Grazzini:2013mca}. The event generation at 13\TeV is tuned so that
the \pt distribution agrees between {\POWHEG2.0} and \textsc{HRes}2.1.
Associated VH production is generated using {\PYTHIA6.4}  ({\PYTHIA8.1}) at  7 and 8 (13)\TeV and normalised to an inclusive cross section
calculated at NNLO QCD and NLO EW precision~\cite{Heinemeyer:2013tqa}.
The expected contribution from $\Pg\Pg\to\Z\PH$ production is estimated using events generated with {\POWHEG2.0} interfaced with {\PYTHIA8.1}.
All signal processes are generated assuming a Higgs boson
mass of 125\GeV, consistent with the combined ATLAS and CMS measurement of the Higgs boson mass~\cite{Aad:2015zhl}.
The SM Higgs boson cross sections at 125\GeV and their uncertainties for
all production mechanisms are taken from Ref.~\cite{YR4} at all centre-of-mass energies. A summary of the simulation used for the different
signal processes is given in Table~\ref{tab:signalmc}.

\begin{table}[hbt]
\topcaption{Simulations used for the different Higgs boson production processes in the 7, 8 and 13\TeV analyses. The \pt
distribution of the ggH production is modified in the 8 TeV simulation to match that predicted with \textsc{HRes} as described in the text.
The accuracy of the inclusive cross section
used for each process is shown, details of which can be found in the text.}
\label{tab:signalmc}
\centering
\cmsTable{ \begin{tabular}{lr|c|c|c|c}
\multicolumn{2}{l}{Production process} & incl. cross section precision& 7\TeV 			  & 8\TeV 			& 13\TeV \\
\hline
\hline
ggH 	& 			       &	N$^3$LO (QCD), NLO (EW)  &	{\POWHEG1.0}+{\PYTHIA6.4} & {\POWHEG1.0}+{\PYTHIA6.4}   & {\POWHEG2.0}+{\PYTHIA8.1} \\
\hline
qqH 	& 			       &	NNLO (QCD), NLO (EW)  &	{\POWHEG1.0}+{\PYTHIA6.4} & {\POWHEG1.0}+{\PYTHIA6.4}   & {\POWHEG2.0}+{\PYTHIA8.1} \\
\hline
\multirow{2}{*}{VH} 	&   $\Pq\Pq\to\mathrm{V}\PH$ &	NNLO (QCD), NLO (EW)  &	{\PYTHIA6.4} 	          & {\PYTHIA6.4}   	        & {\PYTHIA8.1} \\
			&   $\Pg\Pg\to\Z\PH$ &	NNLO (QCD), NLO (EW)  &	{\POWHEG2.0}+{\PYTHIA8.1} & {\POWHEG2.0}+{\PYTHIA8.1}   & {\POWHEG2.0}+{\PYTHIA8.1} \\
\end{tabular}}
\end{table}

The majority of background samples, including $\wjets$, $\zjets$, \ttbar, and
triboson production, are generated using {\MADGRAPH5.1}~\cite{Alwall:2011uj} ({\MADGRAPH5\_aMC@NLO2.2}~\cite{Alwall:2014hca})
with leading order (LO) precision, interfaced with {\PYTHIA6.4} ({\PYTHIA8.1})  for  hadronisation and
fragmentation in the 7 and 8\,(13)\TeV analyses.
Single top quark event samples are produced using {\POWHEG1.0}~\cite{Alioli:2009je} and diboson samples are
generated using {\PYTHIA6.4} ({\PYTHIA8.1}) at 7 and 8 (13)\TeV. QCD multijet events are generated using either {\PYTHIA6.4} or {\MADGRAPH5\_aMC@NLO2.2}, depending
on the analysis. All signal and background samples use the {CTEQ6L}~\cite{Pumplin:2002vw} ( NNPDF3.0~\cite{Ball:2011mu})
parton distribution functions (PDFs) at 7 and 8\,(13)\TeV. The underlying event simulation is done using parameters from the Z2* tune~\cite{Chatrchyan:2013gfi,Khachatryan:2015pea} and the CUETP8M1 tune~\cite{Khachatryan:2015pea} for {\PYTHIA6.4} and {\PYTHIA8.1}, respectively.

The interactions of all final-state particles with the CMS detector are simulated with
{\GEANTfour} \cite{Agostinelli:2002hh}. The simulated samples include PU interactions with
the multiplicity of reconstructed primary vertices matching that in the relevant data sets. An uncertainty of 5\% in the total inelastic pp cross section is propagated to the
PU distribution and is treated as correlated between the data-taking periods.

\section{Analyses included in the combination}
\label{sec:channels}

The characteristic signature of invisible Higgs boson decays for all of the included searches
is a large $\ETm$, with the jets or leptons recoiling against the \ptvecmiss, consistent with one
of the production topologies. In order to reduce the contributions expected from the
SM backgrounds, the properties of the visible recoiling system are exploited. The events are divided into
several exclusive categories designed to target a particular production mode.
A summary of the analyses included in the combination and the
expected signal composition in each of them are given in Table~\ref{tab:analysissummary}.
The VBF search at 8\TeV used in this paper improves on the previous analysis~\cite{Chatrchyan:2014tja} by using
additional data samples from high-rate triggers installed in CMS in 2012.
These triggers wrote data to a special stream, and the events were
reconstructed during the long shutdown of the LHC in 2013~\cite{parked}.
The limit setting procedure has also been updated to allow for a common approach between the 8 and 13\TeV analyses.
The $\zll$ search at 7 and 8\TeV is identical to the one described in Ref.~\cite{Chatrchyan:2014tja} but is described in this paper to
allow for comparison to the 13\TeV analysis.
Both the $\vjj$ and monojet analyses at 8\TeV are re-interpretations of a generic search for
DM production described in Ref.~\cite{Khachatryan:2016mdm} with minor modifications to the selection of events and limit extraction procedure.
In addition to the channels described in the following sections, an 8\TeV analysis targeting ZH
production in which the Z boson decays to a \bbbar pair, described in
Ref.~\cite{Chatrchyan:2014tja}, is included in this combination.

The signal in the VBF analysis is expected to be dominated by qqH production and the expected signals in the $\zll$ and $\zbb$
analyses are composed entirely of ZH production. In contrast, the $\vjj$ and monojet analyses,
which target events with a central, Lorentz-boosted jet, contain a mixture of the different
production modes. This is due to the limited discrimination power of the jet identification used to categorise these events.
As shown in Table~\ref{tab:analysissummary}, the signal composition is similar across the 7 or 8, and 13\TeV data sets.
In the $\vjj$ analysis the ZH contribution is larger, relative to the WH contribution, in the 13\TeV analysis compared to the 8\TeV analysis.
This is because the lepton veto requirement is less efficient at removing leptonic Z boson decays
in the case where the lepton pair is produced at high Lorentz boost causing the isolation cones of the two leptons to
overlap more often at a centre-of-mass energy of 13\TeV compared to 8\TeV.
Each analysis has been optimised separately for the specific conditions and integrated luminosity of the 7, 8, and 13\TeV
data sets leading to differences in the kinematic requirements across the data sets.
These differences are discussed in the following sections.

\begin{table}[hbt]
\topcaption{
Summary of the expected composition of production modes
of a Higgs boson with a mass of 125\GeV
in each analysis included in the combination. The relative
contributions assume SM production cross sections.}
 \label{tab:analysissummary}
   \centering
  \cmsTable{ \begin{tabular}{llccc|cc}
   Analysis 	& Final state        & \multicolumn{3}{c|}{Int. $\mathcal{L}$ ($\fbinv$)}  &\multicolumn{2}{c}{Expected signal composition (\%)}\\

\cline{3-5}\cline{6-7}
   		&		           &  7\TeV & 8\TeV & 13\TeV   &  {7 or 8\TeV} & {13\TeV} \\
   \hline
   qqH-tagged 	& VBF jets  	           & \NA & 19.2~\cite{Chatrchyan:2014tja} & 2.3  & 7.8 (ggH), 92.2 (qqH) & 9.1 (ggH),  90.9 (qqH) \\
   \hline
\multirow{4}{*}{VH-tagged}  & 	\zll       &  4.9~\cite{Chatrchyan:2014tja} & 19.7~\cite{Chatrchyan:2014tja} & 2.3     &  \multicolumn{2}{c}{100 (ZH)} \\
   		& $\zbb$	           & \NA & 18.9~\cite{Chatrchyan:2014tja} & \NA   	  					               &  \multicolumn{2}{c}{100 (ZH)} \\
   		& \multirow{ 2}{*}{$\vjj$} &  \multirow{ 2}{*}{\NA} & \multirow{ 2}{*}{19.7~\cite{Khachatryan:2016mdm}} & \multirow{ 2}{*}{2.3}   & 25.1 (ggH),  5.1 (qqH),   & 38.7 (ggH),  7.1 (qqH), \\
   		& 		           &     & &		  &  23.0 (ZH),  46.8 (WH)    & 21.3 (ZH),  32.9 (WH)   \\
   \hline
   \multirow{ 2}{*}{ggH-tagged}  & \multirow{ 2}{*}{Monojet}		   & \multirow{ 2}{*}{\NA} &\multirow{ 2}{*}{19.7~\cite{Khachatryan:2016mdm}} & \multirow{ 2}{*}{2.3}  & 70.4 (ggH),  20.4 (qqH),  & 69.3 (ggH),  21.9 (qqH), \\
   		& 		           &     & &		  & 3.5 (ZH),   5.7 (WH) 	   & 4.2 (ZH),  4.6 (WH) \\
   \hline
   \end{tabular}}
\end{table}

\subsection{The VBF analysis}

The qqH Higgs boson production mode is characterised by the presence of
two jets with a large separation in $\eta$ and a large invariant mass ($\mjj$).
The selection of events targeting qqH production exploits this distinctive topology
to give good discrimination between the invisible decays of a Higgs boson and the large SM
backgrounds. The contributions from the dominant $\Zvvjets$ and $\Wlvjets$ backgrounds and the QCD
multijet backgrounds are estimated using control regions in data. A simultaneous fit to the yields in the
signal and control regions is performed to extract any potential signal and place upper limits on \brinv.

\subsubsection{Event selection}

{\tolerance=800
Events are selected online using a dedicated VBF trigger, in both the
8 and 13\TeV data sets, with thresholds optimised for the
instantaneous luminosities during each data-taking period. The trigger requires a forward-backward
pair of jets with a pseudorapidity separation of
$\abs{\detajj} > 3.5$ and a large invariant mass. For the majority of the 8\TeV data-taking period
the thresholds used were $\ptjj > 30$ or 35\GeV, depending on the LHC conditions, and $\mjj>700$\GeV.
For the 13\TeV data set, these were modified to $\pt>40$\GeV and $\mjj>600$\GeV. In addition,
the trigger requires the presence of missing transverse energy, reconstructed using the
ECAL and HCAL information only. The thresholds were
$\ETm>40$\,(140)\GeV at 8 (13)\TeV.
The efficiency of the trigger was measured as a
function of the main selection variables: $\ptjj$, $\mjj$, and $\ETm$.
A parameterisation of this efficiency is then applied as a weight to simulated events.
The subsequent selection after the full reconstruction is designed to maintain a trigger efficiency of
greater than 80\%.
\par}

The selection of events  is optimised for VBF production of the Higgs boson
with a mass of 125\GeV, decaying to invisible particles. Events are required to
contain at least two jets within $\abs{\eta}<4.7$
with pseudorapidities of opposite sign, separated by $\abs{\detajj}>3.6$.
The two jets in the event with the highest \pt satisfying this requirement
form the dijet pair. The leading and subleading jets in this pair are required to have
$\pt^{\mathrm{j}_{1}} > 50\,(80)$\GeV, $\pt^{\mathrm{j}_{2}} >45\,(70)$\GeV, and dijet invariant mass
$\mjj>1200\,(1100)$\GeV at 8\,(13)\TeV. Events are required to have $\ETm>90$\,(200)\GeV at 8\,(13)\TeV.

For the 8\TeV dataset, an additional requirement is set on an approximate missing transverse energy
significance variable $S(\ETm)$ defined as the ratio of $\ETm$ to the square root of the scalar sum
of the transverse energy of all PF objects in the event~\cite{Khachatryan:2014gga}. Selected events
are required to satisfy $S(\ETm) > 4\sqrt{\GeVns{}}$.

In order to reduce the large backgrounds from QCD multijet production, the jets in the event are required
to be recoiling against the $\ptvecmiss$. The azimuthal angle
between $\ptvecmiss$ and each jet in the event, $\dphijm$,
is determined. The minimum value of this angle min$\dphijm$ is required to be greater than 2.3.
Finally, events containing at least one muon or electron with $\pt>10$\GeV
are rejected to suppress backgrounds from leptonic decays of the vector boson.

\begin{figure}[hbt]
  \centering
    \includegraphics[width=1.2\cmsFigWidth]{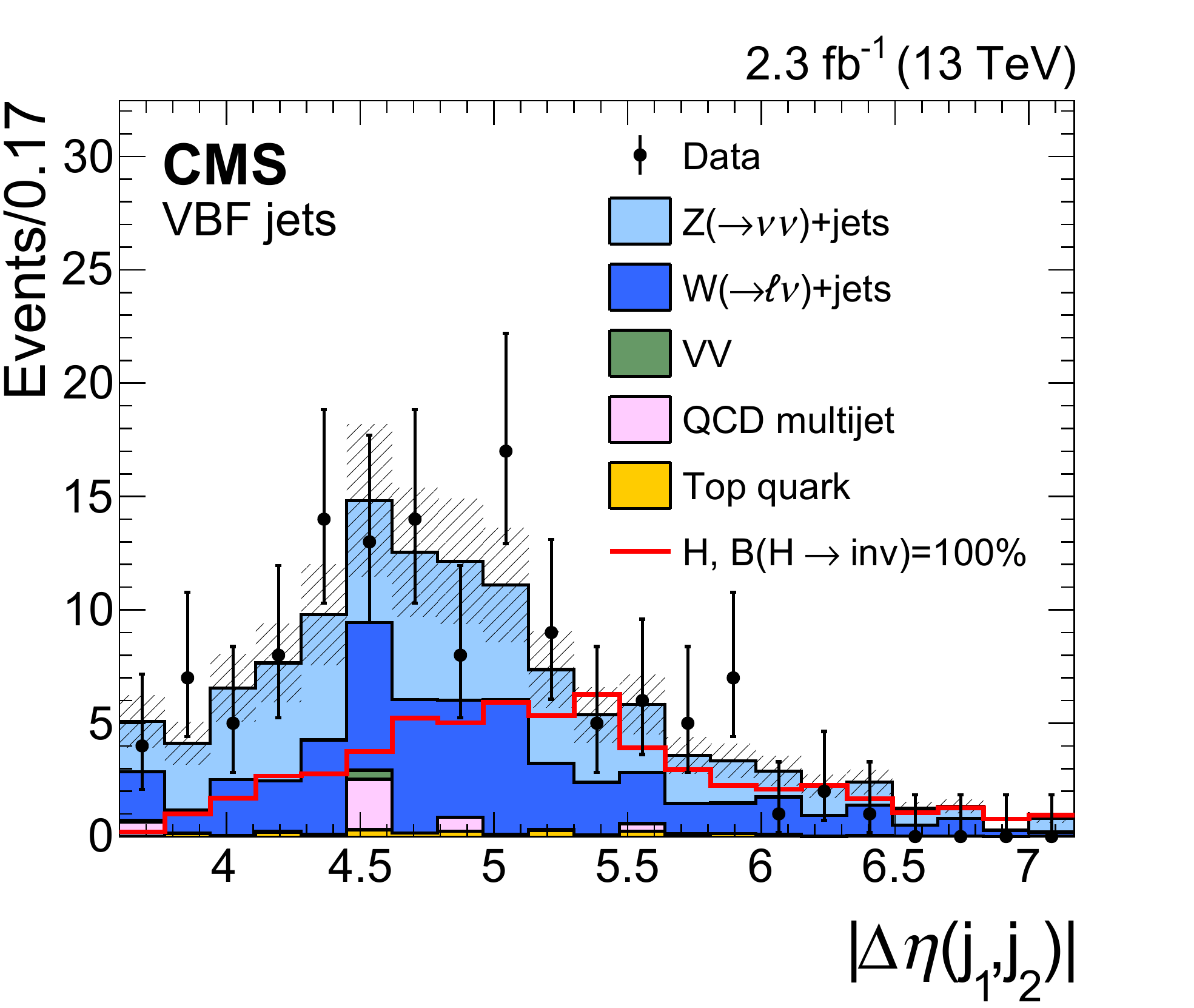} \includegraphics[width=1.2\cmsFigWidth]{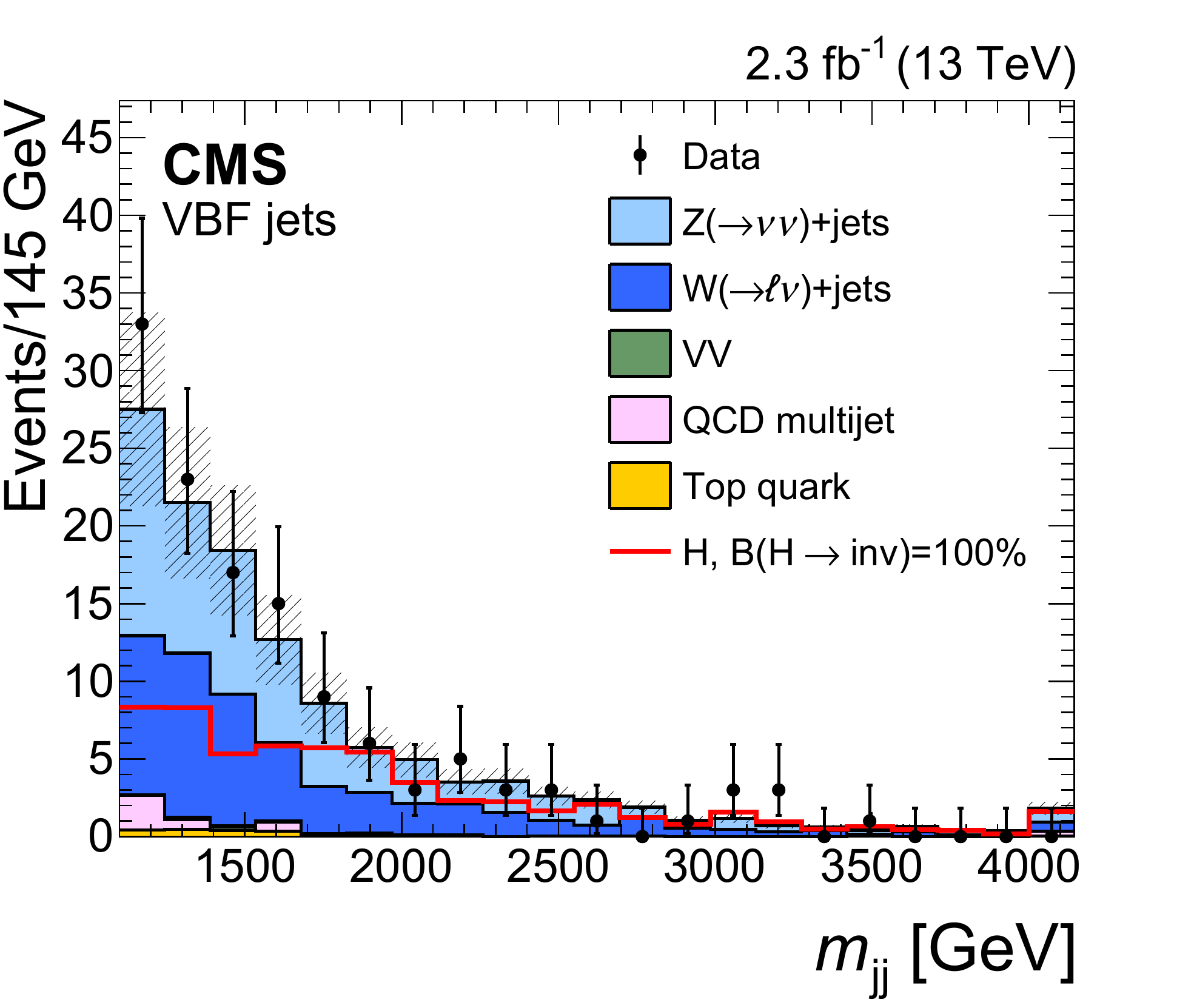}
    \caption{Distributions of (left) $\detajj$ and (right) $\mjj$ in events selected in the VBF analysis
    for data and simulation at 13\TeV.
    The background yields are scaled to their post-fit values, with the total post-fit uncertainty
    represented as the black hatched area. The last bin contains the overflow events. The expected
    contribution from a Higgs boson with a mass of 125\GeV, produced with the SM cross section and decaying
    to invisible particles with 100\% branching fraction, is overlaid.
    }
    \label{fig:vbf_plots}
\end{figure}

A summary of the event selection used in the 8 and 13\TeV data sets is given in
Table~\ref{tab:vbf_event_selection}. Figure~\ref{fig:vbf_plots} shows the distribution of $\detajj$ and
$\mjj$ in data and the predicted background contributions after the selection. The contribution
expected from a Higgs boson with a mass of 125\GeV, produced assuming SM cross sections and decaying
to invisible particles with 100\% branching fraction, is also shown. The backgrounds have been normalised
using the results of a simultaneous fit, as described in Section~\ref{sec:vbf_backgrounds_estimation}.

\begin{table}[h!]
\topcaption{Event selections for the VBF invisible Higgs boson decay search
at 8 and 13\TeV.}
 \label{tab:vbf_event_selection}
 \centering
 \begin{tabular}{lc c}
 \multicolumn{1}{l}{}
    & 8\TeV  &13\TeV \\
 \hline
  $\pt^{\mathrm{j_{1}}}$ & $>$50\GeV & $>$80\GeV   \\
  $\pt^{\mathrm{j_{2}}}$ & $>$45\GeV & $>$70\GeV   \\
  $\mjj$     & $>$1200\GeV & $>$1100\GeV \\
  $\ETm$
  & $>$90\GeV  & $>$200\GeV \\
  $S(\ETm)$ & ${>}4\sqrt{\GeVns{}}$ & \NA\\
  $\mindphi$     & \multicolumn{2}{c}{$>$2.3}    \\
  $\detajj$  & \multicolumn{2}{c}{$>$3.6} \\ \hline
 \end{tabular}
\end{table}

\subsubsection{Background estimation}
\label{sec:vbf_backgrounds_estimation}

The dominant backgrounds to this search arise from
$\Zvvjets$ events and $\Wlvjets$ events with the charged lepton outside of the detector acceptance
or not identified. These backgrounds are estimated using data control regions,
in which a Z or W boson, produced in association with the same dijet topology, decays to
well-identified charged leptons. These control regions are designed to be
as similar to the signal region as possible to limit the extrapolation
required between different kinematic phase spaces. An additional control region,
enriched in QCD multijet events, is defined to estimate the contribution arising
due to mismeasured jet energies causing apparent $\ETm$. Additional smaller contributions due to diboson,
\ttbar, and single top quark production are estimated directly from simulation.

A dimuon control region is defined, enriched in $\zmm$ events,
requiring a pair of oppositely charged muons with $\pt>20$\GeV, $\abs{\eta} <2.1$,
and an invariant mass $m_{\mu\mu}$ in the range  60--120\GeV.
Three single-lepton regions (one enriched in each of the $\wenu$,
$\wmunu$, and $\wtaunu$ processes) are defined by removing the
lepton veto and requiring exactly one isolated lepton, with
$\pt>20$\GeV, of a given flavour, and no leptons of any other flavour. The
lepton is required to be within
$\abs{\eta}<2.1,\,2.4$, or 2.3 for the single-muon, single-electron, or single $\tau$ lepton region,
respectively.  The remaining jets and $\ETm$ criteria are identical
to the signal region, except in the $\wtaunu$
control region where the $\mindphi$ criterion is relaxed to
$\mindphi>1$, taking the minimum over the leading two jets only,
to ensure QCD multijet events are suppressed, while retaining a sufficient number of events in the control region.
Additionally, a requirement that $\mindphi<2.3$ is applied to maintain
an orthogonal selection to the signal region.

Finally, additional control regions are defined in data that are identical
to the signal region selection except for the requirement on $\mindphi$.
In the 8\TeV analysis, a two-step procedure is used in which two control regions are defined.
The first control region is defined by $\mindphi<1$ and is used to determine the distribution of $S(\ETm)$ for QCD multijet events
once the contributions from other backgrounds are subtracted.
The distribution is normalised
using events in a second region defined as $3<S(\ETm)<4\sqrt{\GeVns{}}$ and $1<\mindphi<2$, where the signal contribution is expected to be negligible.
The integral of the normalised distribution in the region $S(\ETm)>4\sqrt{\GeVns{}}$ provides the
estimate of the QCD multijet event contribution in the signal region.
In the 13\TeV analysis, an independent control region is defined by a requirement of $\mindphi < 0.5$
to enrich the QCD multijet contribution.
Systematic uncertainties of 80\% and 100\%
are included at 8 and 13\TeV to account for potential biases in the
extrapolation to the signal region.

Several sources of experimental systematic uncertainties are included
in the predictions of the background components. The dominant ones are
the jet energy scale and resolution~\cite{jec} uncertainties,
which are also propagated to the calculation of the \ETm, resulting
in uncertainties of up to 8\% in the expected background yields. Smaller
uncertainties are included to account for the PU description and lepton reconstruction
efficiencies. Due to the looser selection applied in the $\wtaunu$ control region compared to the signal
region, an additional systematic uncertainty of 20\% in the prediction of the $\wtaunu$
contribution is included.
Finally, additional cross section uncertainties of 7\%\,(10\%)~\cite{tagkey2015250,Khachatryan:2015sga,Chatrchyan:2013oev,Khachatryan:2016txa,Khachatryan:2016tgp}
for diboson production and 10\% (20\%)~\cite{Khachatryan:2016yzq,Khachatryan:2014iya,Khachatryan:2015uqb} for the
top quark background at 8 (13)\TeV are included.

In order to estimate the background contributions, a maximum likelihood fit is performed simultaneously
across each of the control regions, taking the expected background yields from simulation and observed event counts as inputs to the fit.
Two scale factors are included as free parameters in the fit, one scaling both the $\wjets$ and $\zjets$
processes and one scaling the QCD multijet yields across all of the regions.
The fit is thereby able to constrain the contributions from $\wjets$, $\zjets$, and QCD multijets
directly from data.

The ratio of $\Wlvjets$ to $\Zvvjets$ is calculated using simulated samples, generated at LO.
Separate samples are produced for the production of the jets through
quark--gluon vertices (QCD) and production through quark--vector-boson vertices (EW).
A theoretical systematic uncertainty in the expected ratio of the $\Wlvjets$ to
$\Zvvjets$ yields is
derived by comparing LO and NLO predictions after applying the full
VBF kinematic selection using events generated with
{\MADGRAPH5\_a\MCATNLO2.2} interfaced
with {\PYTHIA8.1}, excluding events produced via VBF.
A difference of 30\%
is observed between the ratios predicted by the LO and NLO calculations and is included as a systematic
uncertainty in the ratio of the $\wjets$ to $\zjets$ contributions.
The ratio of the production cross sections of $\Wlvjets$ to $\Zvvjets$ through
EW vertices is compared at NLO and LO precision
using \textsc{vbf@nlo}2.7~\cite{Arnold:2011wj,Baglio:2014uba} and found to
agree within the 30\% systematic uncertainty assigned.

The observed yields in data for each of the control regions in the 13\TeV data set, and the
expected contributions from the backgrounds after the fit ignoring the signal region events,
are given in Table~\ref{tab:yields_vbf_13TeV}.

\begin{table}[hbt]
\topcaption{Post-fit yields for the control regions and
signal region of the VBF analysis using the 13\TeV data set.
The fit ignores the constraints due to the data in the signal region.
For the W and Z processes, jet production through
QCD or EW vertices are listed as separate
entries. The signal yields shown assume SM ggH and qqH production
rates for a Higgs boson with a mass of 125\GeV, decaying to invisible particles
with $\brinv=100\%$.
\label{tab:yields_vbf_13TeV}}
 \centering
 \begin{tabular}{ll|c|c|c|c|c|c}
   \multicolumn{2}{c}{Process}&  \multicolumn{1}{c}{Signal} & \multicolumn{5}{c}{Control regions} \\
\cline{1-2} \cline{4-8}
  &\multicolumn{1}{c}{} & \multicolumn{1}{c}{Region} & Single e & Single $\mu$ & Single $\tau$ & $\mpmm $ & QCD  \\
 \hline

 \multirow{2}{*}{$\Zmmjets$}
& QCD& \NA & \NA & \NA & \NA & $4.2 \pm 1.1$ & \NA \\
& EW & \NA & \NA & \NA & \NA & $2.0 \pm 0.7$ & \NA \\
 \multirow{2}{*}{$\Zvvjets$}
& QCD& $47 \pm 12$     & \NA & \NA & \NA & \NA & \NA \\
& EW& $21 \pm 7$      & \NA & \NA & \NA & \NA & \NA \\
 \multirow{2}{*}{$\Wmvjets$}
& QCD& $13 \pm 2$      & \NA & $53 \pm 5$ & $0.4 \pm 0.2$ & \NA & $45 \pm 5$ \\
& EW& $4.3 \pm 0.8$   & \NA & $27 \pm 3$ & \NA & \NA & $6.0 \pm 0.9$ \\
 \multirow{2}{*}{$\Wevjets$}
& QCD& $9.3 \pm 1.5$   & $17 \pm 3$ & \NA & $0.2 \pm 2.2$ & \NA & $39 \pm 4$\\
& EW& $5.4 \pm 1.1$   & $7.8 \pm 1.3$ & \NA & $0.2 \pm 0.1$ & \NA & $6.1 \pm 1.0$ \\
 \multirow{2}{*}{$\Wtvjets$}
& QCD& $13 \pm 2$      & $0.06 \pm 0.06$ & \NA & $12 \pm 2$ & \NA & $74 \pm 9$ \\
& EW& $5.5 \pm 1.2$   & \NA & \NA & $5.1 \pm 1.2$ & \NA & $24 \pm 3$ \\
Top quark  &          & $2.3 \pm 0.4$   & $1.5 \pm 0.3$ & $6.8 \pm 0.9$ & $7.1 \pm 1.0$ & $0.22 \pm 0.06$ & $82 \pm 11$ \\
QCD multijet   &          & $3   \pm 23$    & \NA & $5 \pm 3$ & $0.4 \pm 0.3$ & \NA & $1200 \pm 170$ \\
Dibosons       &          & $0.7 \pm 0.3$   & $0.4 \pm 0.4$ & $0.8 \pm 0.4$ & \NA & $0.02 \pm 0.02$ & $1.8 \pm 0.7$ \\
 \hline
Total bkg. &          & $125 \pm 28$    & $27 \pm 3$ & $91 \pm 8$ & $25 \pm 4$ & $6.4 \pm 1.4$ & $1500 \pm 170$ \\
 \hline
Data       &          & $126$ & $29$ & $89$ & $24$ & $7$ & $1461$
\\
 \hline
Signal 			  & qqH &\multicolumn{1}{c}{ $53.6 \pm 4.9$} &  \multicolumn{5}{c}{}\\
$m_{\PH}=125$\GeV  & {ggH} & \multicolumn{1}{c}{$5.4  \pm 3.6$} &  \multicolumn{5}{c}{}\\
 \cline{1-3}
 \end{tabular}
\end{table}

\subsection{The \texorpdfstring{$\zll$}{ZllH} analysis}

The ZH production mode, where the Z boson decays to a pair of charged leptons, has a smaller
cross section than qqH but a clean final state with lower background.
The search targets events with a pair of same-flavour, opposite-charge leptons ($l=\Pe,~\mu$),
consistent with a leptonic Z boson decay, produced in association
with a large $\ETm$. The background is dominated by the diboson processes, $\zzllvv$ and
$\wzlvll$, which contribute roughly 70\% and 25\% of the total background, respectively.

In the 7 and 13\TeV data sets the sensitivity of the search is enhanced by using the
distribution of the transverse mass of the dilepton-$\ETm$ system $\mt$, defined as
\begin{linenomath*}
\begin{equation*}
\mt =  \sqrt{2\ptll\ETmiss\left[1 - \cos{\Delta\phi(\ell\ell,\ptvecmiss)}\right]},
\end{equation*}
\end{linenomath*}
where $\ptll$ is the transverse momentum of the dilepton system and
$\Delta\phi(\ell\ell,\ptvecmiss)$ is the azimuthal angle between the dilepton system and
the missing transverse momentum vector. In the 8\TeV data set, a two-dimensional fit is performed to the distributions
of $\mt$ and the azimuthal angle between the two leptons $\Delta\phi(\ell,\ell)$ to
exploit the increased statistical precision available in that data set~\cite{Chatrchyan:2014tja}.

\subsubsection{Event selection}

Events for this channel are recorded using double-electron and double-muon triggers,
with thresholds of $\pt^{\Pe} > 17\,(12)$\GeV and $\pt^{\mu} > 17\,(8)$\GeV at
13\TeV and $\pt^{\Pe,\mu} > 17\,(8)$\GeV at 7 and 8\TeV, for the leading (subleading)
electron or muon, respectively. Single-electron and single-muon
triggers are also included in order to recover residual trigger inefficiencies.

Selected events are required to have two well-identified, isolated leptons with the
same flavour and opposite charge ($\epem$ or $\mpmm$), each
with $\pt > 20$\GeV,  and an invariant mass within the range 76--106\GeV.
In the 13\TeV analysis, the $\dyll$ background is substantially suppressed by requiring $\Delta\phi(\ell,\ell)<\pi/2$.
As little hadronic activity is expected in the $\zllh$ channel, events
with more than one jet with $\pt>30$\GeV are rejected.
Events containing a muon with $\pt > 3$\GeV and a b jet with $\pt>30$\GeV
are vetoed to reduce backgrounds from top quark production.
Diboson backgrounds are suppressed by rejecting events containing additional electrons or muons
with $\pt > 10$\GeV. In the 13\TeV analysis, events containing a $\tau$ lepton with $\pt>20$\GeV are
vetoed to suppress the contributions from WZ production.

The remainder of the selection has been optimised for a Higgs boson with a mass of 125\GeV,
produced in the $\zllh$ production mode. As a result of this
optimisation, events are required to have $\ETm > 120\,(100)$\GeV, $\Delta\phi(\ell\ell,\ptvecmiss)>2.7$
(2.8), and $|\ETm-\ptll|/\ptll < 0.25\,(0.4)$, in the 7 and 8 (13)\TeV data sets.
Finally, the events are required to have $\mt>$ 200\GeV.
A summary of the event selection used for the 7, 8, and 13\TeV data sets is given in
Table~\ref{tab:zll_event_selection}.

\begin{table}[h!]
	\topcaption{Event selections for the $\zll$ invisible Higgs boson search
	using the 7, 8, and 13\TeV data sets. The $\dphijm$ requirement is applied only in the 1-jet category.
 \label{tab:zll_event_selection}}
 \centering
 \begin{tabular}{l|c c}
 \multicolumn{1}{l|}{}	    & \multicolumn{1}{c}{7 and 8\TeV}  & 13\TeV  \\
 \hline
  $\pt^{\Pe,\mu}$     	     & \multicolumn{2}{c}{${>}20$\GeV} \\
  $\mll$     		     & \multicolumn{2}{c}{76--106\GeV} \\
  $\Delta\phi(\ell,\ell)$ 	      & \NA   & ${<}\pi/2$          \\
  $\ETm$     		      & ${>}120$\GeV  & ${>}100$\GeV        \\
  $\Delta\phi(\ell\ell,\ptvecmiss)$ & ${>}2.7$      & ${>}2.8$            \\
  $\dphijm$      	      & \NA          & ${>}0.5$  	         \\
  $|\ETm-\ptll|/\ptll$        & ${<}0.25$     & ${<}0.4$            \\
  $\mt$     		     & \multicolumn{2}{c}{$>$200\GeV} \\
 \hline
 \end{tabular}
\end{table}

The selected events are separated into two categories, events that contain no jets with $\pt>30$\GeV and
$\abs{\eta} < 4.7$, and events that contain exactly one such jet. An additional selection requiring
$\dphijm>0.5$ is applied in the 1-jet category at 13\TeV which significantly reduces the contribution
from $\zjets$ events.

\begin{figure}[hbt]
  \centering
 \includegraphics[width=1.2\cmsFigWidth]{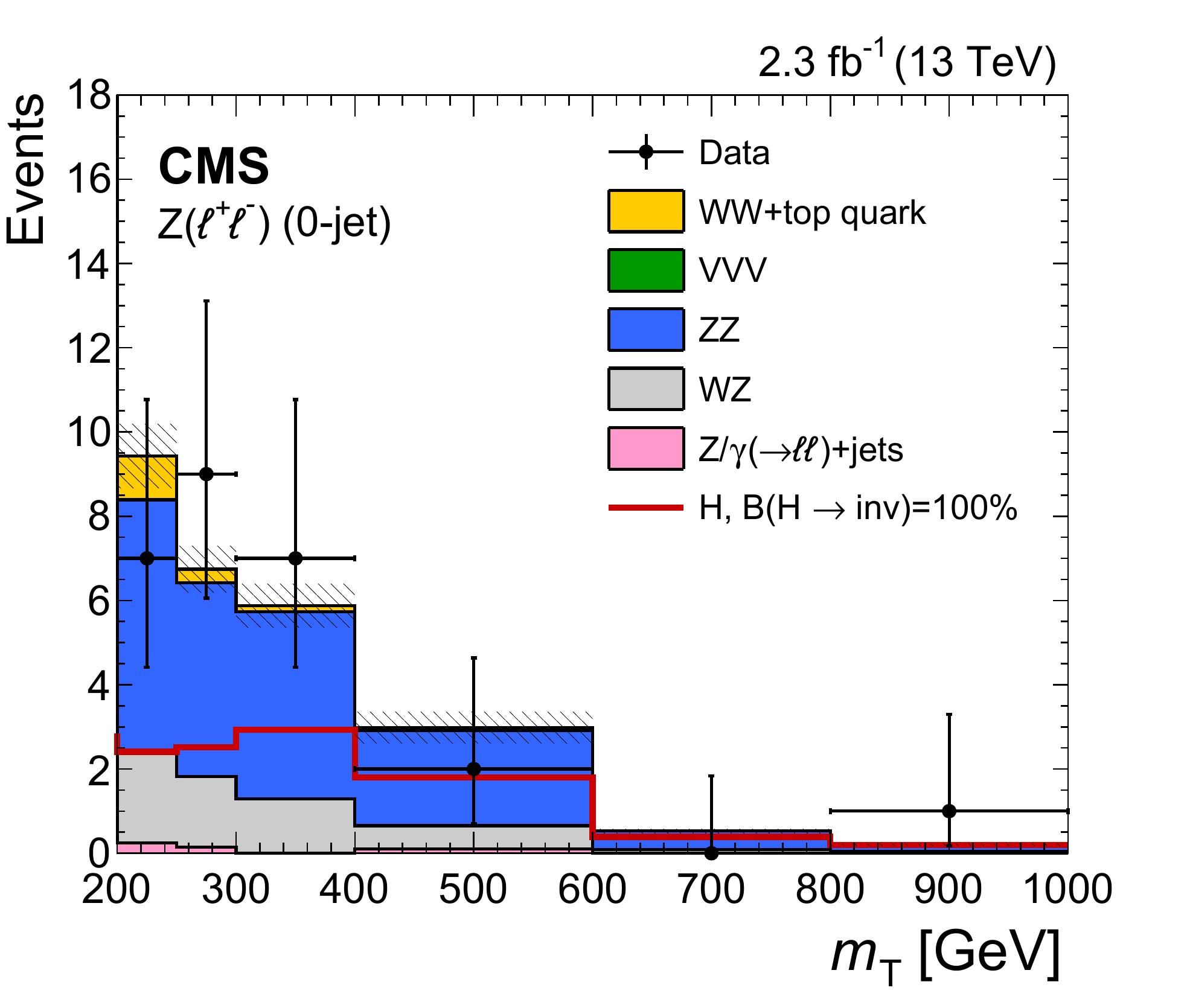}
\includegraphics[width=1.2\cmsFigWidth]{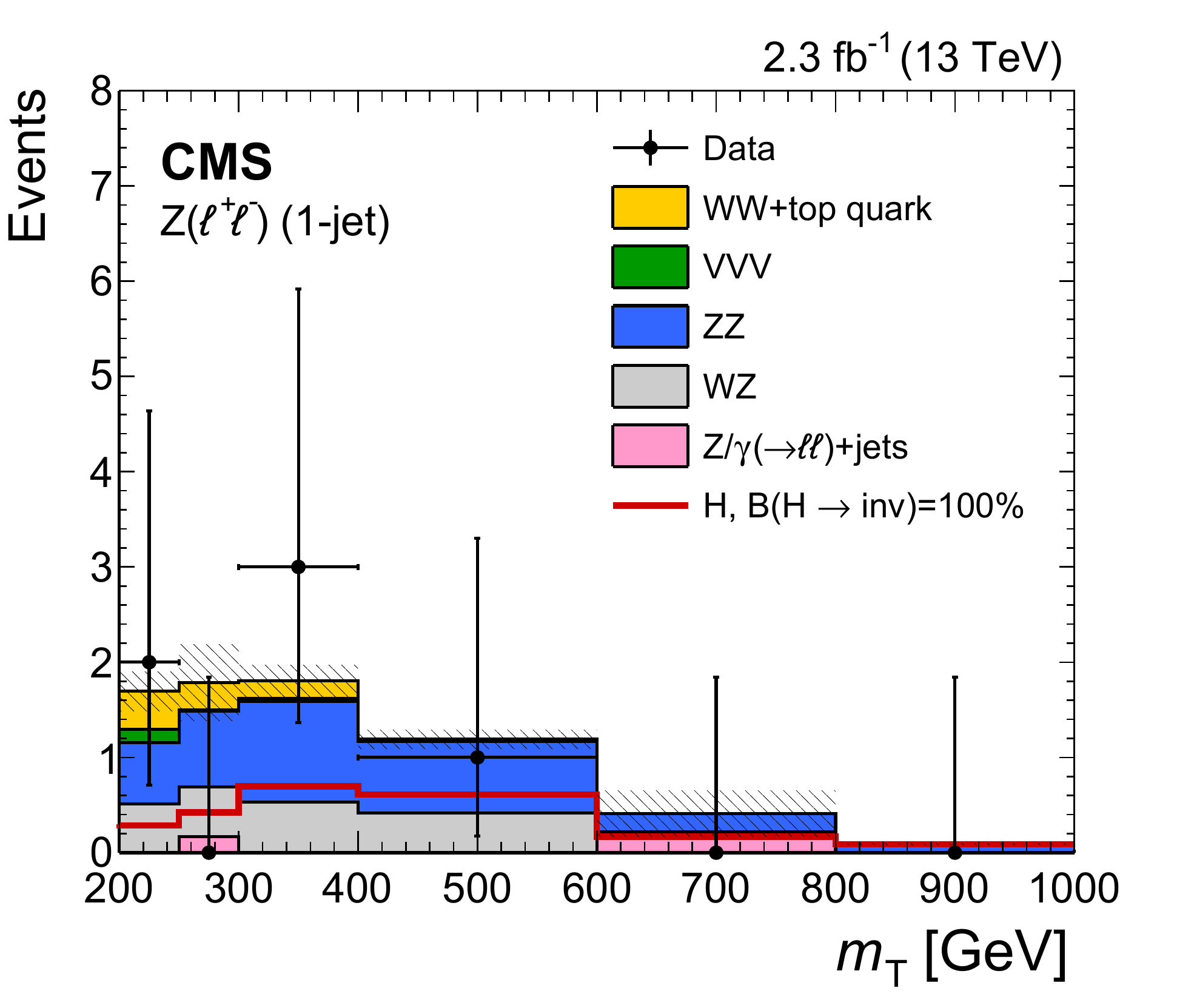}
    \caption{Distributions of $\mt$ in data and simulation for events
    in the (left) 0-jet and (right) 1-jet categories of the $\zll$ analysis at 13\TeV, combining dielectron and dimuon events.
    The background yields are normalised to 2.3\fbinv. The shaded bands
    represent the total statistical and systematic uncertainties in the backgrounds.
    The horizontal bars on the data points represent the width of the bin centred at that point.
    The expectation from a Higgs boson with a mass of 125\GeV, from
    ZH production, decaying to invisible particles with a 100\% branching fraction is shown in red.
    \label{fig:zll_plots}}
\end{figure}

The distributions of $\mt$ for selected events in data and simulation, combining electron and muon events,
for the 0-jet and 1-jet categories at 13\TeV are shown in Fig.~\ref{fig:zll_plots}.

\subsubsection{Background estimation}
The dominant backgrounds, \zzllvv~and \wzlvll,~are generated at NLO using {\POWHEG2.0},
for production via $\mathrm{q}\bar{\mathrm{q}}$. Corrections are applied to account for
higher-order QCD and EW effects which are roughly 10--15\% each but with opposite sign.
The contribution from $\textrm{gg}\to\mathrm{ZZ}$ is estimated using \textsc{mcfm7.0}~\cite{Campbell:2011bn}.
Uncertainties due to missing higher-order corrections for these processes are evaluated by varying
the renormalisation and factorisation scales up and down by a factor of two, yielding systematic
uncertainties between 4 and 10\%. A 2\% uncertainty is added to account for the jet category
migration due to uncertainties in the PDFs used in the signal generation, calculated following the
procedures outlined in Ref.~\cite{Butterworth:2015oua}.
Additional uncertainties are included in the $\qqbar\to \Z\Z$ event yield
to account for the uncertainties in the higher-order corrections applied.

The $\zjets$ background is estimated using a data control region dominated by
single-photon production in association with jets ($\phojets$). The $\phojets$ events
have similar jet kinematics to \DYjets, but with a much larger production rate.
The $\phojets$ events are weighted, as a
function of the photon \pt, to match the distribution observed in $\DYjets$ events in
data. This accounts for the dependence of the $\ETm$  on the hadronic activity.
A systematic uncertainty of 100\% is included in the final $\zjets$ background estimate to account for the
limited number of events at large \pt in the data used to weight the $\phojets$ events.

The remaining, nonresonant backgrounds are estimated using a control
sample selecting pairs of leptons of different flavour and opposite charge
(${\Pe^{\pm}\mu^{\mp}}$) that pass all of the signal region selections. These
backgrounds consist mainly of leptonic W boson decays in
\ttbar and $\PQt\PW$ processes, and $\PW\PW$ events. Additionally, leptonic
$\tau$ lepton decays contribute to these backgrounds.
As the branching fraction to the ${\Pe^\pm\PGm^\mp}$ final
states is twice that of the $\Pep\Pem$ or $\PGmp\PGmm$ final states, the
${\Pe^\pm\PGm^\mp}$ control region provides precise estimates of the nonresonant backgrounds.
In the 13\TeV analysis, the contribution from the nonresonant backgrounds is given by
\begin{linenomath*}
\begin{equation*}
N_{\ell\ell}^\text{bkg}= N_{\Pe\mu}^\text{data} (k_{\Pe\Pe/\mu\mu} + 1/k_{\Pe\Pe/\mu\mu})/2,
\end{equation*}
\end{linenomath*}
where $N_{\Pe\mu}^\text{data}$ is the number of events in the
${\Pe^\pm\mu^\mp}$ control region after subtracting other backgrounds and
$k_{\Pe\Pe/\mu\mu}=\sqrt{\smash[b]{N_{\Pe\Pe}/N_{\mu\mu}}}$ is a correction factor accounting
for the differences in acceptance and efficiency for electrons and muons,
measured using $\dyee$ and $\dymm$ events in data. An uncertainty of 70\% in the
estimated yield of the nonresonant backgrounds is included to account for
the statistical and systematic uncertainties of the extrapolation from the
$\Pe^{\pm}\mu^{\mp}$ control region. A similar method using sideband regions around the
Z boson mass peak was used to estimate these backgrounds in the 8\TeV analysis,
as described in Ref.~\cite{Chatrchyan:2014tja}. This method was also used in the
13\TeV analysis as a cross check and the differences between the results of the two methods
of 10--15\% are included as additional systematic uncertainties.

Additional uncertainties in the background estimates arise from uncertainties in the lepton efficiencies, momentum
scale, jet energy scale and resolution, and \ETm energy scale and resolution. Each of these contributes around
2\% uncertainty in the normalisation of the dominant backgrounds. Statistical uncertainties are included for all simulated samples.
These uncertainties are
propagated as both shape and normalisation variations of the predicted $\mt$ distributions.

The numbers of expected and observed events for the 0-jet
and 1-jet categories in the 13\TeV analysis are given in Table~\ref{tab:zll_yeilds}.
The signal yield assumes the SM ZH production rate for a Higgs
boson with a mass of 125\GeV decaying to invisible particles with 100\% branching fraction.

\begin{table}[h!]
 \topcaption{Predicted signal and background yields and
 observed number of events after full selection in the 13\TeV $\zll$-tagged analysis. The numbers are given
 for the 0-jet and 1-jet categories, separately for the $\epem$
 and $\mpmm$ final states. The
 uncertainties include statistical and systematic components.
 The signal prediction assumes a SM ZH production rate for a Higgs boson with the
 mass of 125\GeV and a 100\% branching fraction to invisible particles.
 \label{tab:zll_yeilds}
 }
 \centering
 \begin{tabular}{l|cc|cc}
\multicolumn{1}{c}{Process} & \multicolumn{2}{c|}{0 jets}                  & \multicolumn{2}{c}{1 jet}             \\
\cline{2-5}
 \multicolumn{1}{c}{}                         		& $\mpmm$     & $\epem$        & $\mpmm$              & $\epem$               \\
\hline
$\mathrm{ZH}$, $\mathrm{m_H}=125$\GeV  &  5.97 $\pm$ 0.55    &4.27 $\pm$ 0.39    &1.29 $\pm$ 0.20 & 0.98 $\pm$ 0.15  \\
\hline
$\DYjets$                   &    0.45 $\pm$ 0.45  &0.30 $\pm$ 0.30    &0.45 $\pm$ 0.45 & 0.30 $\pm$ 0.30\\
$\zzllvv$       			&    10.4 $\pm$ 1.14   &7.46 $\pm$ 0.81   &2.04 $\pm$ 0.31 & 1.49 $\pm$ 0.23\\
$\wzlvll$   				&    3.42 $\pm$ 0.28  &2.40 $\pm$ 0.19    &1.04 $\pm$ 0.10 & 1.00 $\pm$ 0.10\\
Top/WW/$\tau\tau$     			&    0.69 $\pm$ 0.23  &0.88 $\pm$ 0.29    &0.44 $\pm$ 0.22 & 0.26 $\pm$ 0.13\\
VVV                       		&     	   \NA         &      \NA           &0.13 $\pm$ 0.06 & 0.07 $\pm$ 0.03\\
\hline
Total background          		&    15.0 $\pm$ 1.28    & 11.0 $\pm$ 0.93  &4.10 $\pm$ 0.60 & 3.12 $\pm$ 0.41 \\
\hline
Data                      		&  18              & 8              &  5            	&  1		  \\
\hline
 \end{tabular}
\end{table}

\subsection{The \texorpdfstring{$\vjj$}{VJJ} and monojet analyses}

Searches for final states with central jets and $\ETm$ suffer from large backgrounds. However, the ggH mode and the
VH associated mode, in which the vector boson decays hadronically, have relatively large signal contributions
despite the tight requirements on the jets. The search strategies for the VH mode, in which the vector boson decays hadronically, and ggH modes are
very similar, targeting events with large $\ETm$, with the $\ptvecmiss$ recoiling against jets from either
gluon radiation or a hadronically
decaying vector boson. Events are divided into two categories, depending on the jet properties.
The dominant backgrounds arise from $\Zvvjets$ and $\Wlvjets$ events, accounting for 90\% of the
total background. These backgrounds are estimated using control regions in data
and a simultaneous fit to the $\ETm$ distribution of the events across all regions is performed to extract a potential signal.

\subsubsection{Event selection}

The data set is collected using a suite of triggers with requirements on $\ETm$ and hadronic activity.
In the 8\TeV analysis two triggers are used: the first
requires $\ETm>120$\GeV, while the second requires $\ETm>95$ or 105\GeV,
depending on the data-taking period, together with a jet of $\pt>80$\GeV and $\abs{\eta}<2.6$.
In the 13\TeV data set, the trigger requires $\ETm>90$\GeV and $H_{\mathrm{T}}^{\text{miss}} > 90$\GeV,
where $H_{\mathrm{T}}^{\text{miss}}$ is defined as the magnitude of the vector sum of the \pt
of all jets with $\pt>20$\GeV.
In both 8 and 13\TeV data sets the
calculation of $\ETm$  does not include muons, allowing for the same triggers to be used in the signal,
single-muon and dimuon control regions. For events selected for the analysis, the trigger efficiency is
found to be greater than 99\%\,(98\%) at 8\,(13)\TeV.

To reduce the QCD multijet background the events in the 8\TeV analysis that do not
satisfy the requirement that the angle between the $\ptvecmiss$
and the leading jet $\Delta\phi(\ptvecmiss,\mathrm{j}) > 2$ are removed. In the 13\TeV data set the
requirement is instead $\mindphi>0.5$, where the minimum is over the four leading jets  in the event.
Events in the signal regions of the 8 (13)\TeV
analysis are vetoed if they contain an electron or muon with $\pt>10$\GeV,
a photon with $\pt > 10\,(15)$\GeV, or a $\tau$ lepton with $\pt>18\,(15)$\GeV.
Backgrounds from top quark decays are suppressed by applying a veto on
events containing a b jet with $\pt> 15$\GeV.

Selected events are classified by the topology of the jets in order to distinguish initial- or final-state radiation
from hadronic vector boson decays. This results in two exclusive event categories to target two channels: the monojet and $\vjj$.
If the vector boson decays hadronically and has sufficiently high \pt, its hadronic decay products
are captured by a single reconstructed large-radius jet. Events in the $\vjj$ channel are required to have $\ETm>250$\GeV
and contain a reconstructed $R= 0.8$ jet with $\pt>200\,(250)$\GeV and $\abs{\eta}<2.0\,(2.4)$ in the  8\,(13)\TeV analysis.
Additional requirements are included to better identify jets from the decay of a vector boson by using the ``subjettiness''
quantity $\tau_2/\tau_1$, as defined in Refs.~\cite{Thaler:2010tr,Thaler:2011gf}, which identifies jets with a two subjet
topology, and the pruned jet mass ($m_{\text{prune}}$)~\cite{Ellis:2009me}.
The $\tau_2/\tau_1$ ratio is required to be smaller than 0.5 (0.6) and $m_{\text{prune}}$
is required to be in the range  60--110\,(65--105)\GeV in the 8\,(13)\TeV analysis. The optimisation of the selection for VH production is performed
independently for the 8 and 13\TeV data sets.

If an event fails the $\vjj$ selection, it can instead be included in the monojet channel.
Events in this channel are required to contain at least one anti-\kt jet, reconstructed with cone size 0.5\,(0.4),
with $\pt>150$ (100)\GeV and $\abs{\eta}<2.0$ (2.5) in the 8 (13)\TeV analysis. In the 8\TeV analysis, only events with up to two jets are
included in the $\vjj$ and monojet categories, provided that the separation of the second
jet from the leading jet in azimuthal angle satisfies $\Delta\phi<2$. For the purposes of this requirement, only jets
reconstructed with the anti-\kt algorithm using a cone size of 0.5 are counted beyond the leading jet in the $\vjj$ channel.
This requirement on the maximum number of jets $N_\mathrm{j}$ was dropped for the 13\TeV analysis to increase the signal acceptance.
Finally, events are required to have $\ETm>200$\GeV.

A summary of the event selection for the $\vjj$ and monojet categories
is given in Table~\ref{tab:ggh_vjj_selection}.
In addition to this selection, events that pass the corresponding VBF selection are vetoed to avoid an overlap with the VBF search.

\begin{table}[h!]
	\caption{Event selections for the $\vjj$ and monojet
	invisible Higgs boson decay searches using the 8 and 13\TeV data sets. The requirements
	on $\pt^{\mathrm{j}}$ and $\abs{\eta}^{\mathrm{j}}$ refer to the highest \pt (large-radius) jet in the monojet ($\vjj$) events.
	The 8\TeV analysis uses only the leading jet in the definition of $\mindphi$. In the 8\TeV number of jets $N_\mathrm{j}$ selection,
	events with one additional jet are allowed if this additional jet falls within $\Delta\phi$ of the leading jet as described in the text.}
 \label{tab:ggh_vjj_selection}
 \centering
 \begin{tabular}{l|c c|c c}
 \multicolumn{1}{l}{}	    & \multicolumn{2}{c|}{8\TeV}  &\multicolumn{2}{c}{13\TeV}\\ \cline{2-5}
 \multicolumn{1}{l}{}	    & $\vjj$ & Monojet        & $\vjj$ & Monojet     \\
 \hline
  $\pt^{\mathrm{j}}$     	    & ${>}200$\GeV & ${>}150$\GeV &  ${>}250$\GeV & ${>}100$\GeV  \\
  $\abs{\eta}^{\mathrm{j}}$     	    & \multicolumn{2}{c|}{${<}2$}  &  ${<}2.4$	  & ${<}2.5$	\\
  $\ETm$     		    	    & ${>}250$\GeV & ${>}200$\GeV  &  ${>}250$\GeV & ${>}200$\GeV \\
  $\tau_2/\tau_1$      		    & ${<}0.5$ 	   & \NA 	        & ${<}0.6$     & \NA  \\
  $m_{\text{prune}}$      	    & 60--110\GeV  & \NA 	        & 65--105\GeV     & \NA  \\
  $\mindphi$      	    &\multicolumn{2}{c|}{${>}2$ rad}& \multicolumn{2}{c}{${>}0.5$ rad} \\
  $N_\mathrm{j}$  	    & \multicolumn{2}{c|}{$=$1} &\multicolumn{2}{c}{\NA} 	\\
 \hline
 \end{tabular}
\end{table}

\subsubsection{Background estimation}

The dominant $\Zvvjets$~and $\Wlvjets$~backgrounds are estimated from control regions in
data consisting of dimuon, single-muon, and \phojets~events. In the 13\TeV analysis,
additional control regions consisting of dielectron and single-electron events are used.
The $\ETm$ in each control region is redefined to mimic the $\ETm$ distribution of the
$\Zvvjets$~and $\Wlvjets$ backgrounds in the signal region by excluding the leptons or the
photon from the computation of $\ETm$.

A dimuon control region is defined by selecting events that contain two opposite-sign muons with
$\pt^{\mu_{1},\mu_{2}} > 10\,(20), 10$\GeV at 8 (13)\TeV and an invariant mass between 60 and 120\GeV. A
single-muon control region is defined by selecting events with an isolated muon with $\pt>20$\GeV.

A dielectron control region in the 13\TeV data is defined using similar requirements on the
two electrons as for the dimuon control region. Single-electron triggers with a \pt threshold of 27\GeV are used to record the events,
and at least one of the selected electrons, after the full event reconstruction, is required to have $\pt>40$\GeV.
Additionally a single-photon trigger with a \pt threshold of 165\GeV is used to recover events in which the \pt
of the Z boson is large (more than 600\GeV), leading to inefficiencies in the electron isolation requirements.
A single-electron control sample is selected using the same triggers. The \pt of the electron in this region
is required to be greater than 40\GeV in order to reach the region in which the trigger is fully efficient.
An additional requirement of $\ETm > 50$\GeV is imposed on single-electron events in order to
suppress the QCD multijet background.

The use of dilepton events to constrain the $\Zvvjets$ background suffers from large statistical uncertainties since the branching fraction of
the Z boson to neutrinos is roughly six times larger than that to muons or electrons. In order to overcome this, $\phojets$ events are
additionally used to reduce the statistical uncertainty at the cost of introducing theoretical uncertainties in their use for modelling
\Zvvjets~events~\cite{zjetspapers}.
The \phojets~control sample is constructed using single-photon triggers. Events are required to have a well isolated
photon with $\pt>170$ (175)\GeV and $\abs{\eta}<2.5$\,(1.44) in the 8\,(13)\TeV analysis to ensure a $\phojets$ purity
of at least 95\%~\cite{Khachatryan:2015iwa}.

The events in all control regions are required to pass
all of the selection requirements applied in the signal region, except for the lepton and photon vetoes.
As in the signal region, events in the control regions are separated into $\vjj$ and monojet channels.

The $\ETm$ distribution of the $\Zvvjets$~and $\Wlvjets$ backgrounds is estimated from a maximum likelihood fit,
performed simultaneously across all $\ETm$ bins in the signal and control regions. The expected numbers of $\Zvvjets$
(and $\Wlvjets$ in the 8\TeV analysis) in each bin of $\ETm$ are free parameters of the fit.
For each bin in $\ETm$, the ratio of the $\Zvvjets$ yield in the signal region
to the corresponding yields of the $\Zmmjets$, $\Zeejets$ and $\phojets$ processes in
the dimuon, dielectron, and $\phojets$ control regions are used to determine the expectations in these control regions for given values of the fit
parameters~\cite{Khachatryan:2016mdm}. Similarly, the ratio of the $\Wlvjets$ yield in the signal region to the corresponding yields
of the $\Wmvjets$ and $\Wevjets$ processes in the single-muon and single-electron control regions are used to determine the expectations
in these two control regions.
The ratios are determined from simulation after applying \pt-dependent NLO QCD $K$-factors derived using the
{\sc MadGraph5\_aMC@NLO2.2} MC generator and \pt-dependent NLO EW $K$-factors derived from
theoretical calculations~\cite{Kuhn:2005gv,Kallweit:2014xda,Kallweit:2015fta,Kallweit:2015dum}.  In the 8\TeV analysis, the ratio between the two backgrounds is left unconstrained in the fit.
In the 13\TeV analysis, the ratio of $\Wlvjets$ to $\Zvvjets$ in the
signal region is constrained to that predicted in simulation after the application of NLO QCD and EW $K$-factors.

Systematic uncertainties are included to account for theoretical uncertainties in the
$\gamma$ to Z and W to Z differential cross section ratios due to the choice of the
renormalisation and factorisation scales and uncertainties in the PDFs used to generate the events~\cite{Ball:2014uwa}.
The value of the systematic uncertainty in these differential cross sections due to higher-order
EW corrections is taken to be the full NLO EW correction, which can be as
large as 20\% for large values of \ETm.
For the kinematic region in which the $K$-factors are applied, the interference between QCD and EW effects reduces the
correction obtained compared to applying the $K$-factors independently~\cite{Kallweit:2015dum}.
The difference between accounting for this
interference or not is covered by the systematic uncertainties applied.
Uncertainties in the selection efficiencies of muons, electrons,
photons (up to 2\%), and hadronically decaying $\tau$ leptons (3\%) are included. The
uncertainty in the modelling of \ETm in simulation is dominated by the jet energy scale
uncertainty and varies between 2 and 5\%, depending on the \ETm bin.

The remaining subdominant backgrounds due to top quark and diboson processes are estimated
directly from simulation. Systematic uncertainties of 10 and 20\% are included in the cross sections for the top
quark~\cite{Khachatryan:2015uqb} and diboson backgrounds~\cite{Khachatryan:2016txa,Khachatryan:2016tgp}.
An additional 10\% uncertainty is assigned to the top quark backgrounds to account for the discrepancies observed between
data and the simulation in the \pt distribution of the \ttbar pair.
An inefficiency of the $\vjj$ tagging requirements can cause events to migrate between the $\vjj$ and monojet channels.
An uncertainty in the $\vjj$ tagging efficiency of 13\%, which allows for migration of events between the $\vjj$ and
monojet channels, is included to account for this. This uncertainty comprises a statistical component which is
uncorrelated between the 8 and 13\TeV analyses and a systematic component which is fully correlated.

In the 8\TeV data set, the contribution from QCD multijet events is determined using simulation normalised to the data,
while in the 13\TeV data set the contribution is determined using a dedicated control sample.
Although large uncertainties are included to account for the extrapolation from the control region to the
signal region, the impact on the final results is small.

\begin{figure}[hbt]
  \centering
    \includegraphics[width=1.2\cmsFigWidth]{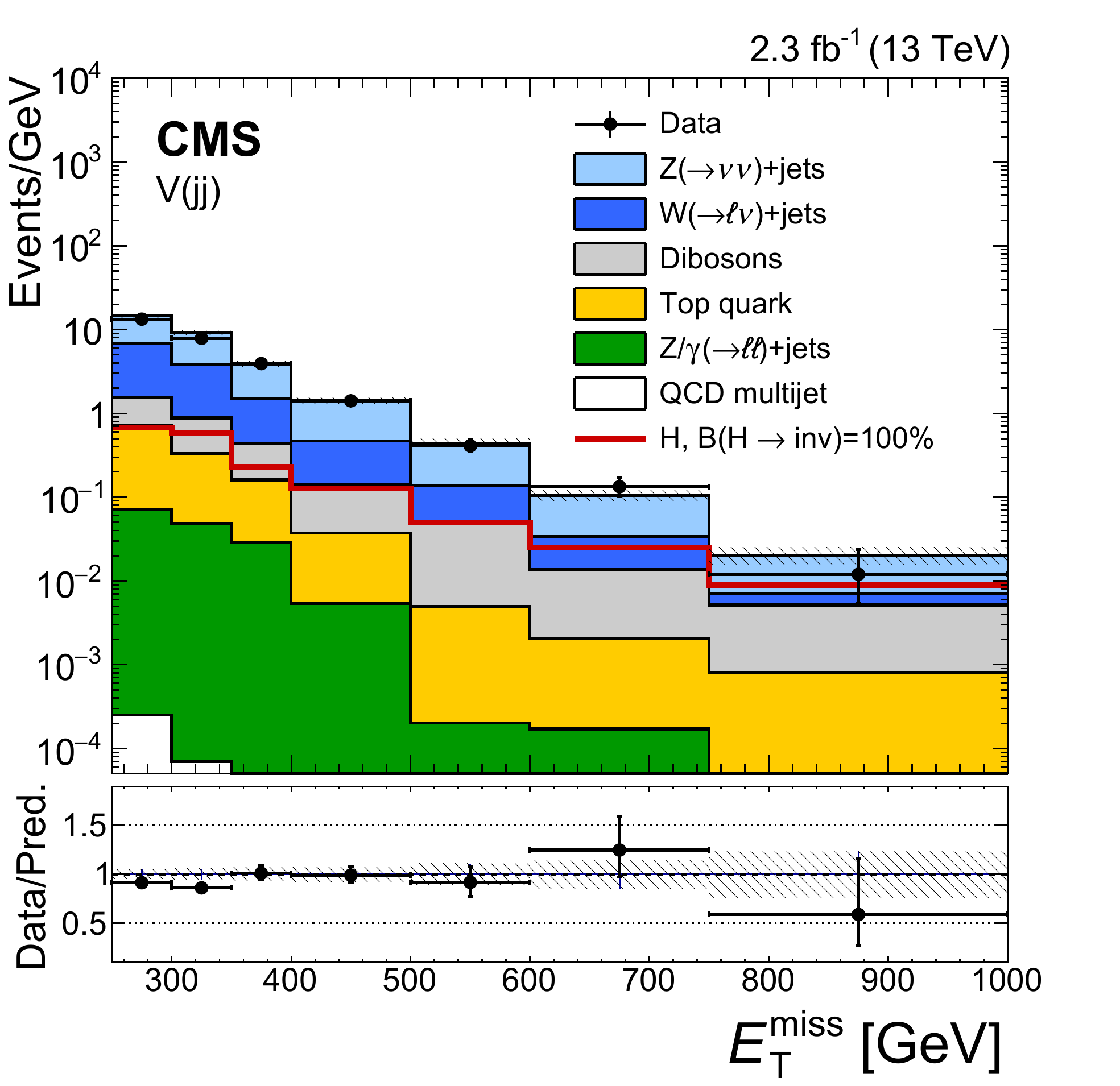}
    \includegraphics[width=1.2\cmsFigWidth]{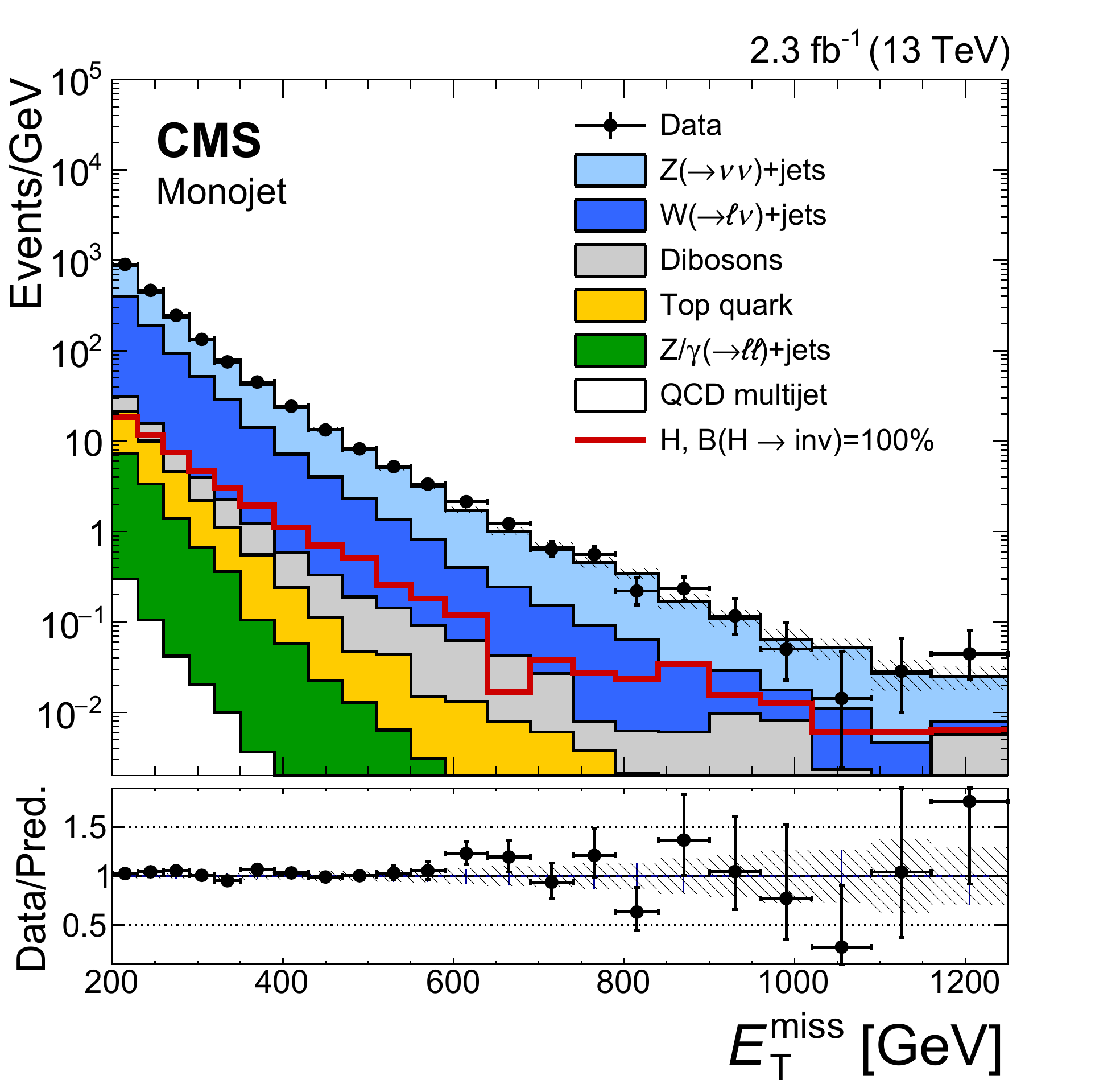}
    \caption{Distributions of $\ETm$ in data and predicted background contributions in the (left) $\vjj$
    and (right) monojet channels at 13\TeV. The background prediction is taken from a fit using only the
    control regions and the shaded bands represent the
    statistical and systematic uncertainties in the backgrounds after that fit.
    The horizontal bars on the data points represent the width of the bin centred at that point.
    The expectations from a Higgs boson with a mass of 125\GeV decaying to invisible particles with a branching fraction of
    100\% are superimposed.
    }
    \label{fig:exo_plots}
\end{figure}

Figure~\ref{fig:exo_plots} shows the distribution of $\ETm$ in data for the $\vjj$ and monojet channels
in the 13\TeV analysis and the background predicted after performing a simultaneous fit, which ignores the constraints from
data in the signal regions. The signal expectation assuming SM rates for production of a Higgs boson with a mass of 125\GeV
with $\brinv=100$\% is superimposed.

\section{Results}
\label{sec:results}

No significant deviations from the SM expectations are observed in any of the searches performed.
The results are interpreted in terms of upper limits on \brinv under various assumptions about the
Higgs boson production cross section, $\sigma$. Limits are calculated using an asymptotic approximation of the
CL$_\mathrm{s}$ prescription~\cite{junkcls,Read1} using a profile likelihood ratio test statistic~\cite{Cowan:2010js},
in which systematic uncertainties are modelled as nuisance parameters $\boldsymbol{\theta}$ following a frequentist approach~\cite{HiggsCombination}.

The profile likelihood ratio is defined as,
\begin{linenomath*}
\begin{equation*}
	q = -2 \ln \frac{ L(\text{data}|\sigbrinv,\hat{\hat{\boldsymbol{\theta}}})}{ L(\text{data}|\sighatbrinv,\hat{\boldsymbol{\theta}})},
\end{equation*}
\end{linenomath*}
where $\sighatbrinv$ represents the value of $\sigbrinv$, which maximises the likelihood $L$
for the data, and $\hat{\boldsymbol{\theta}}$ and $\hat{\hat{\boldsymbol{\theta}}}$ denote the unconditional maximum
likelihood estimates for the nuisance parameters and the estimates for a specific value of $\sigbrinv$.
The value of $\sigbrinv$ is restricted to be positive when maximising the likelihood. The ``data'' here refers to the data
in all of the control and signal regions for each analysis described in Section~\ref{sec:channels}.

The statistical procedure accounts for correlations between the nuisance parameters in each of
the analyses. The uncertainties in the diboson cross sections, the lepton efficiencies, momentum
scales, and the integrated luminosity are correlated across all categories
of a given data set. The uncertainties in the inclusive signal cross sections are additionally correlated
across the measurements at 7, 8, and 13\TeV.

The kinematics of the jets selected in the VBF channel are distinct from those selected in the $\vjj$ and monojet channels.
For this reason, the jet energy scale and resolution uncertainties are considered uncorrelated between those channels.
The b jet energy scale and resolution uncertainties for the $\zbb$ channel are estimated using
a different technique from that used for other jets and so are treated as uncorrelated with other
searches~\cite{Chatrchyan:2013zna}.

Where simulation is used to model the $\ETm$ distributions of the signal or backgrounds, uncertainties are propagated
from the jet and lepton energy scales and resolutions as well as from modelling of the unclustered energy. These uncertainties are treated as
fully correlated between the 7, 8, and 13\TeV  data sets, except for the 8\TeV $\vjj$ and monojet channels for which independent calibrations
based on control samples in data are applied.

Systematic uncertainties in the inclusive ggH, qqH, and VH production cross sections due to renormalisation and factorisation
scales, and PDF uncertainties are taken
directly from Ref.~\cite{YR4} and treated as fully correlated across the 7, 8, and 13\TeV data sets.
An additional systematic uncertainty of 50\% in the ggH production cross section of the Higgs boson in
association with two jets is included for the contribution of ggH production in the
VBF categories. This uncertainty is estimated by comparing the two-jet NLO
generators {\POWHEG2.0+\textsc{minlo}}~\cite{Hamilton:2012np} and
{a\MCATNLO}~\cite{Frixione:2002ik} interfaced with {\HERWIGpp2.3}~\cite{Bahr:2008pv}.
Furthermore, an uncertainty in the Higgs boson \pt distribution in
ggH production is included in the monojet channels and estimated by varying the renormalisation and factorisation
scales~\cite{Catani:2003zt}. This uncertainty is correlated between the 8 and 13\TeV categories.
Uncertainties in the acceptance arising from uncertainties in the PDFs used to determine the expected signal yields are evaluated independently
for the different signal processes in each event category and treated as additional normalisation nuisance parameters.

\subsection{Upper limits on \texorpdfstring{\brinv}{B(H to inv)} assuming SM production}

\begin{figure}[hbt]
  \centering \includegraphics[width=2.\cmsFigWidth]{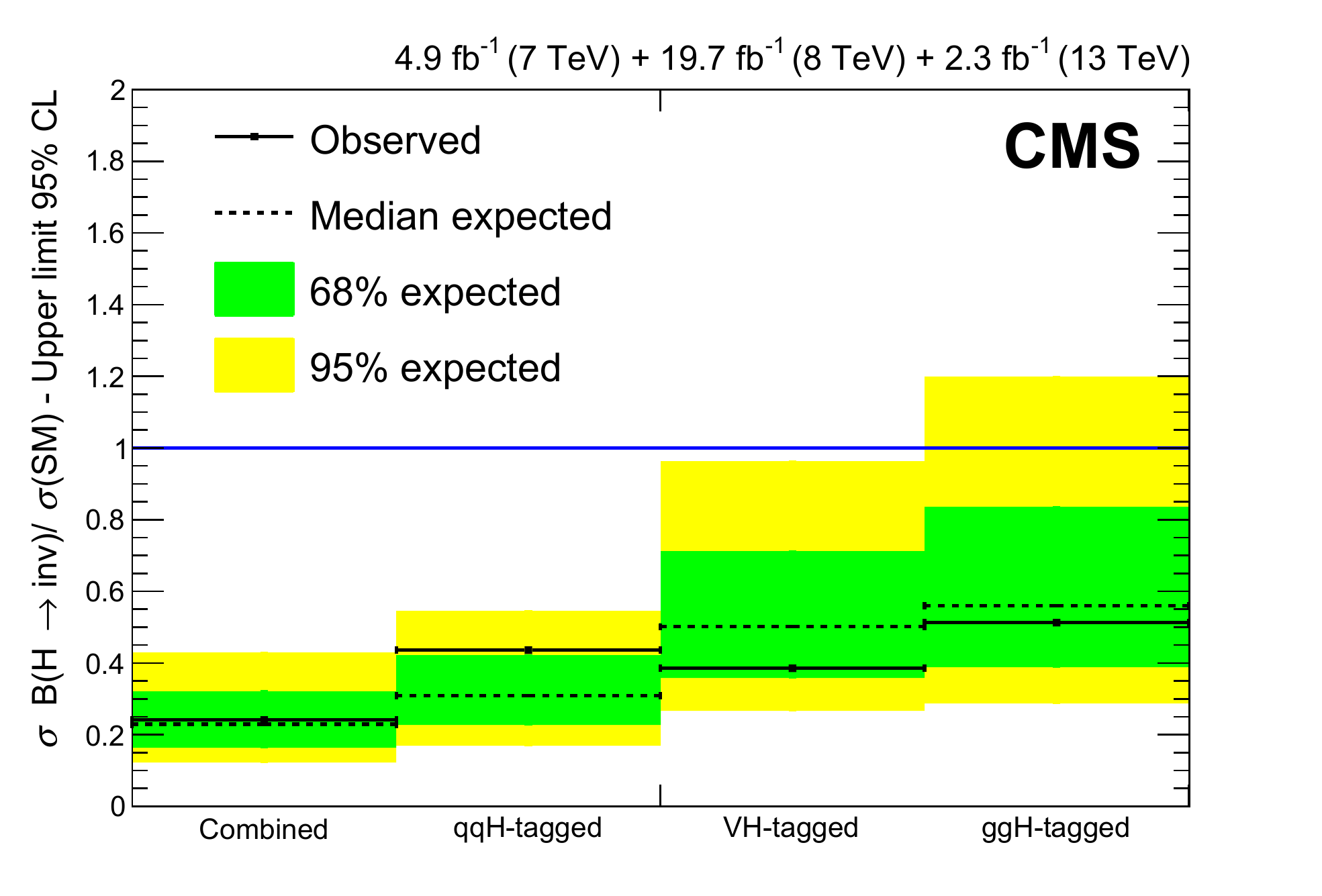}
    \caption{Observed and expected 95\% CL limits on $\sigbrinv$ for individual combinations
    of categories targeting qqH, VH, and ggH production, and the full combination assuming
    a Higgs boson with a mass of 125\GeV.}
    \label{fig:comblimits}
\end{figure}

Observed and expected upper limits on $\sigbrinv$, where $\sigma(\mathrm{SM})$ is the total SM Higgs boson production cross section,
are determined at the 95\% CL and presented in Fig.~\ref{fig:comblimits}.
The limits are obtained from the combination of all categories and from sub-combinations of categories, which target one of the ggH, qqH, and
VH production mechanisms, corresponding to the analysis tags in Table~\ref{tab:analysissummary}.
The relative contributions from the different production mechanisms
in these results are fixed to their SM predictions within the uncertainties.
If the production cross sections take their SM values, the results can be used to constrain
the branching fraction of the Higgs boson to invisible particles. Assuming SM production rates for the ggH, qqH, and VH modes,
the combination yields an observed (expected) upper limit of $\brinv<\higgsbrobs$ $(\higgsbrexp)$ at the 95\% CL.

\begin{figure}[hbt]
  \centering
  \includegraphics[width=1.2\cmsFigWidth]{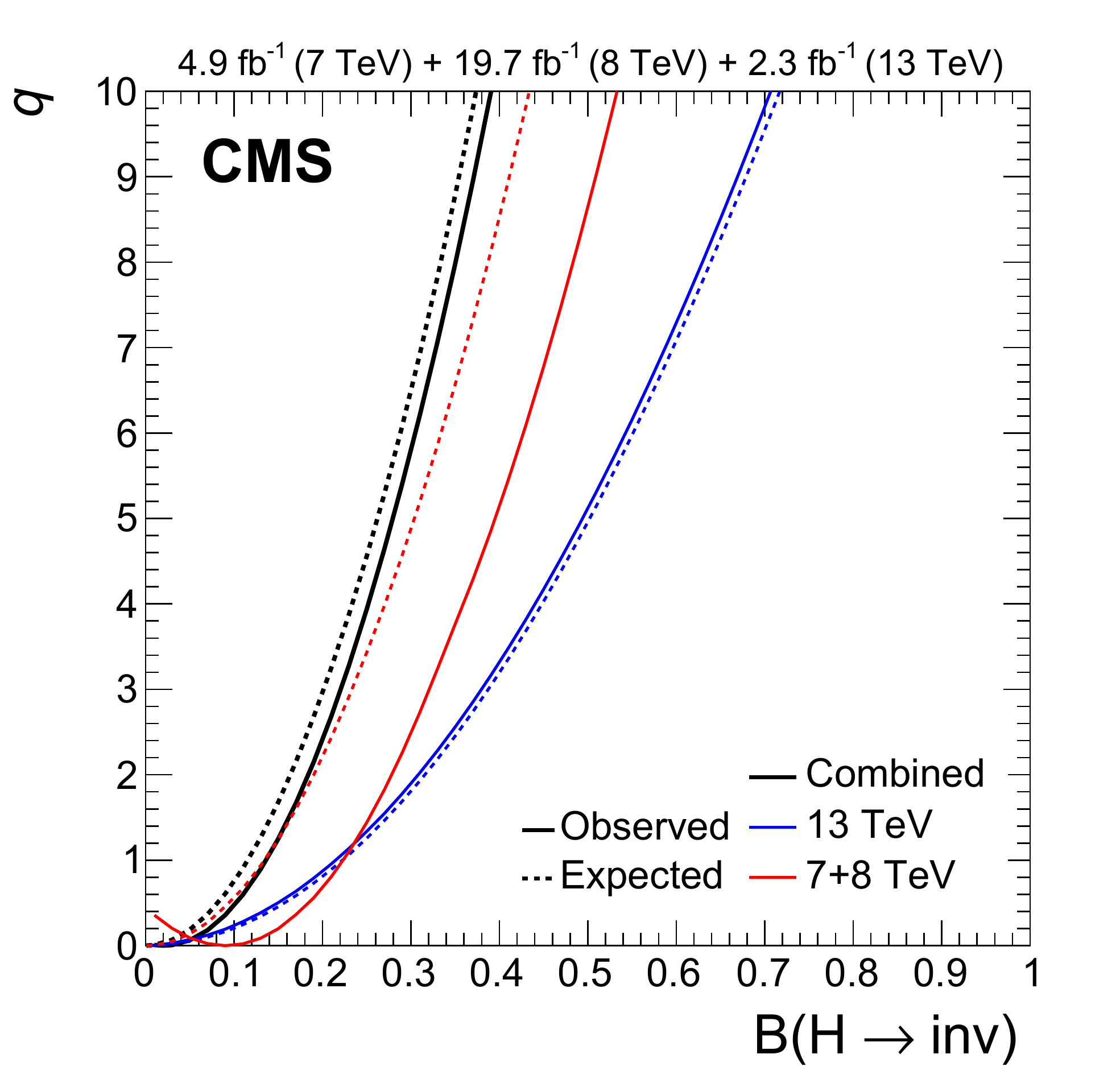}
  \includegraphics[width=1.2\cmsFigWidth]{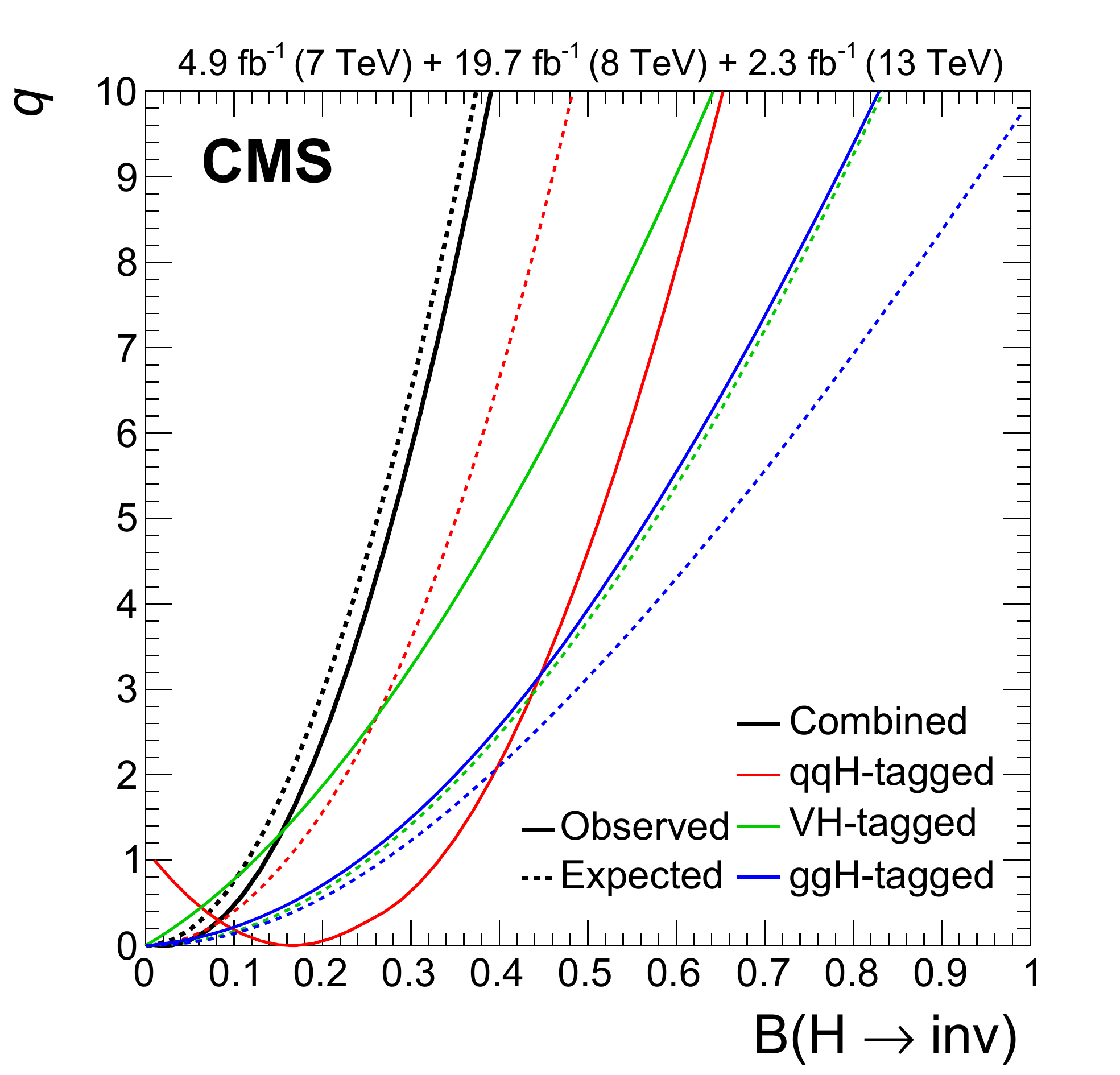}
\caption{Profile likelihood ratio as a function of \brinv assuming SM production
    cross sections of a Higgs boson with a mass of 125\GeV.
    The solid curves represent the observations in data and the
    dashed curves represent the expected result assuming no
    invisible decays of the Higgs boson.
    (left) The observed and expected likelihood scans for the partial combinations
    of the 7+8 and 13\TeV analyses, and the full combination.
    (right) The observed and expected likelihood scans for the partial combinations
    of the qqH-tagged, VH-tagged, and ggH-tagged analyses, and the full combination.
    }
  \label{fig:lhscan}
\end{figure}

The profile likelihood ratios as a function of \brinv using partial combinations of the 7+8 and 13\TeV analyses, and for the
full combination are shown in Fig.~\ref{fig:lhscan}\,(left).
The profile likelihood ratio scans for the partial combinations of the qqH-tagged,
VH-tagged, and ggH-tagged analyses are shown in Fig.~\ref{fig:lhscan}\,(right).
The results are shown for the data and for an Asimov
data set, defined as the data set for which the maximum likelihood estimates of all parameters are equal to their true values~\cite{Cowan:2010js},
in which $\brinv=0$ is assumed.

The dominant systematic uncertainties for the qqH-tagged, $\zll$, $\vjj$, and ggH-tagged searches in the 13\TeV data set are
listed in Tables~\ref{tab:impacts_vbf},~\ref{tab:impacts_zll},~\ref{tab:impacts_vh}, and~\ref{tab:impacts_ggh}, respectively.

The impact of each independent source of systematic uncertainty is calculated for an Asimov data set in which $\sigbrinv$ is assumed to be 1.
The impact is defined as the maximum difference in the fitted value of $\sigbrinv$, when varying the nuisance parameter associated to
that source of systematic uncertainty within one standard deviation of its maximum likelihood estimate. The total systematic uncertainty, and
the total uncertainty fixing all nuisance parameters associated to systematic uncertainties
that are not expected to improve with additional luminosity (statistical only), for each analysis is also shown.
Finally, the total uncertainty is given for each analysis. The statistical only and total uncertainties are determined from the interval in
$\sigbrinv$ for which $q<1$. The total systematic uncertainty is determined by subtracting the statistical only uncertainty from the total
uncertainty in quadrature.
With the luminosity of the 13\TeV data set, the
sensitivity of the qqH-tagged and $\zll$ analyses is dominated by the statistical uncertainty while for the $\vjj$ and ggH-tagged
analyses, a reduction in the theoretical and experimental systematic uncertainties related to the modelling of the \Zvvjets~and \Wlvjets~backgrounds
would yield significant improvements.

\begin{table}[!htbp]
\topcaption{
Dominant sources of systematic uncertainties and their impact on the fitted
value of \brinv in the VBF analysis at 13\TeV.
The systematic uncertainties are split into common uncertainties and those
specific to the signal model.
The total systematic uncertainty, the total uncertainty
fixing all constrained nuisance parameters to their maximum likelihood estimates (statistical only), and the total uncertainty are also given.
}
\label{tab:impacts_vbf}
\centering
\begin{tabular}{lr}
Systematic uncertainty  & Impact\\
\hline
\multicolumn{2}{l}{Common} \\[1.2ex]

W to Z ratio in QCD produced $\vjets$ & 13\%\\
W to Z ratio in EW produced $\vjets$& 6.3\%\\
Jet energy scale and resolution & 6.0\%\\
QCD multijet normalisation & 4.3\%\\
Pileup mismodelling & 4.2\%\\
Lepton efficiencies & 2.5\%\\
Integrated luminosity & 2.2\%\\
\hline
\multicolumn{2}{l}{Signal specific} \\[1.2ex]
ggH acceptance        & 3.8\%\\
Renorm. and fact. scales and PDF (qqH) & 1.8\%\\
Renorm. and fact. scales and PDF (ggH) & ${<}0.2$\%\\[1.2ex]
\hline
\rule{0pt}{3ex}Total systematic       & $_{-19}^{+15}\%$ \\
\rule{0pt}{3ex}Total statistical only & $_{-27}^{+28}\%$ \\
\rule{0pt}{3ex}Total uncertainty      & $_{-33}^{+32}\%$ \\
\hline
\end{tabular}
\end{table}

\begin{table}[!htbp]
\topcaption{
Dominant sources of systematic uncertainties and their impact on the fitted
value of \brinv in the $\zll$ analysis at 13\TeV.
The systematic uncertainties are split into common uncertainties and those
specific to the signal model.
The total systematic uncertainty, the total uncertainty
fixing all constrained nuisance parameters to their maximum likelihood estimates (statistical only), and the total uncertainty are also given.
}
\label{tab:impacts_zll}
\centering
\begin{tabular}{lr}
Systematic uncertainty  & Impact\\
\hline
\multicolumn{2}{l}{Common} \\[1.2ex]
ZZ background, theory & 16\%\\
Integrated luminosity 	     & 8.4\%\\
b tagging efficiency & 6.2\%\\
Electron efficiency & 6.2\%\\
Muon efficiency & 6.2\%\\
Electron energy scale & 3.2\%\\
Muon momentum scale   & 3.2\%\\
Jet energy scale & 2.2\%\\
Diboson normalisation & 5.3\%\\
$\Pe\mu$ region extrapolation & 4.0\%\\
\zll~normalisation & 4.8\%\\
\hline
\multicolumn{2}{l}{Signal specific} \\[1.2ex]
Renorm. and fact. scales and PDF (qqZH) & 7.4\%\\
Renorm. and fact. scales and PDF (ggZH) & 4.0\%\\
\hline
\rule{0pt}{3ex}Total systematic       & $_{-23}^{+27}\%$ \\
\rule{0pt}{3ex}Total statistical only & $_{-50}^{+56}\%$ \\
\rule{0pt}{3ex}Total uncertainty      & $_{-55}^{+62}\%$ \\
\hline
\end{tabular}\end{table}\newpage

\begin{table}[!htbp]
\topcaption{
Dominant sources of systematic uncertainties and their impact on the fitted
value of \brinv in the \vjj analysis at 13\TeV.
The systematic uncertainties are split into common uncertainties and those
specific to the signal model.
The total systematic uncertainty, the total uncertainty
fixing all constrained nuisance parameters to their maximum likelihood estimates (statistical only), and the total uncertainty are also given.
}
\label{tab:impacts_vh}
\centering
\begin{tabular}{lr}
Systematic uncertainty  & Impact\\
\hline
\multicolumn{2}{l}{Common} \\[1.2ex]
$\phojets/\Zvvjets$ ratio, theory & 32\%\\
$\Wlvjets/\Zvvjets$ ratio, theory & 21\%\\
Jet energy scale and resolution & 12\%\\
$\vjj$-tagging efficiency & 12\%\\
Lepton veto efficiency & 13\%\\
Electron efficiency & 13\%\\
Muon efficiency & 8.6\%\\
b tagging efficiency & 5.7\%\\
Photon efficiency & 3.1\%\\
\ETm scale & 4.6\%\\
Top quark background normalisation & 6.0\%\\
Diboson background normalisation & ${<}1$\%\\
Integrated luminosity & ${<}1$\%\\
\hline
\multicolumn{2}{l}{Signal specific} \\[1.2ex]
ggH \pt-spectrum      & 12\%\\
Renorm. and fact. scales and PDF (ggH) & 3.0\%\\
Renorm. and fact. scales and PDF (VH)  & 1.4\%\\
\hline
\rule{0pt}{3ex}Total systematic       & $_{-51}^{+55}\%$ \\
\rule{0pt}{3ex}Total statistical only & $_{-46}^{+50}\%$ \\
\rule{0pt}{3ex}Total uncertainty      & $_{-69}^{+74}\%$ \\
\hline
\end{tabular}\end{table}\newpage

\begin{table}[!htbp]
\topcaption{
Dominant sources of systematic uncertainties and their impact on the fitted
value of \brinv in the monojet analysis at 13\TeV.
The systematic uncertainties are split into common uncertainties and those
specific to the signal model.
The total systematic uncertainty, the total uncertainty
fixing all constrained nuisance parameters to their maximum likelihood estimates (statistical only), and the total uncertainty are also given.
}
\label{tab:impacts_ggh}
\centering
\begin{tabular}{lr}
Systematic uncertainty  & Impact\\
\hline
\multicolumn{2}{l}{Common} \\[1.2ex]
Muon efficiency & 24\%\\
Electron efficiency & 22\%\\
Lepton veto efficiency & 16\%\\
b jet tag efficiency & 3.2\%\\
$\Wlvjets/\Zvvjets$ ratio, theory & 16\%\\
$\phojets/\Zvvjets$ ratio, theory& 5.8\%\\
Jet energy scale and resolution & 10\%\\
\ETm scale & 1.8\%\\
Integrated luminosity & 3.0\%\\
Diboson background normalisation & 2.7\%\\
Top quark background normalisation & ${<}1$\%\\
\hline
\multicolumn{2}{l}{Signal specific} \\[1.2ex]
ggH \pt-spectrum       & 15\%\\
Renorm. and fact. scales and PDF (ggH)  & 5.8\%\\
\hline
\rule{0pt}{3ex}Total systematic       & $_{-50}^{+57}\%$ \\
\rule{0pt}{3ex}Total statistical only & $_{-22}^{+25}\%$ \\
\rule{0pt}{3ex}Total uncertainty      & $_{-55}^{+62}\%$ \\
\hline
\end{tabular}\end{table}\newpage

\subsection{Non-SM production and DM interpretations}

\begin{figure}[hbt]
\centering
  \includegraphics[width=1.5\cmsFigWidth]{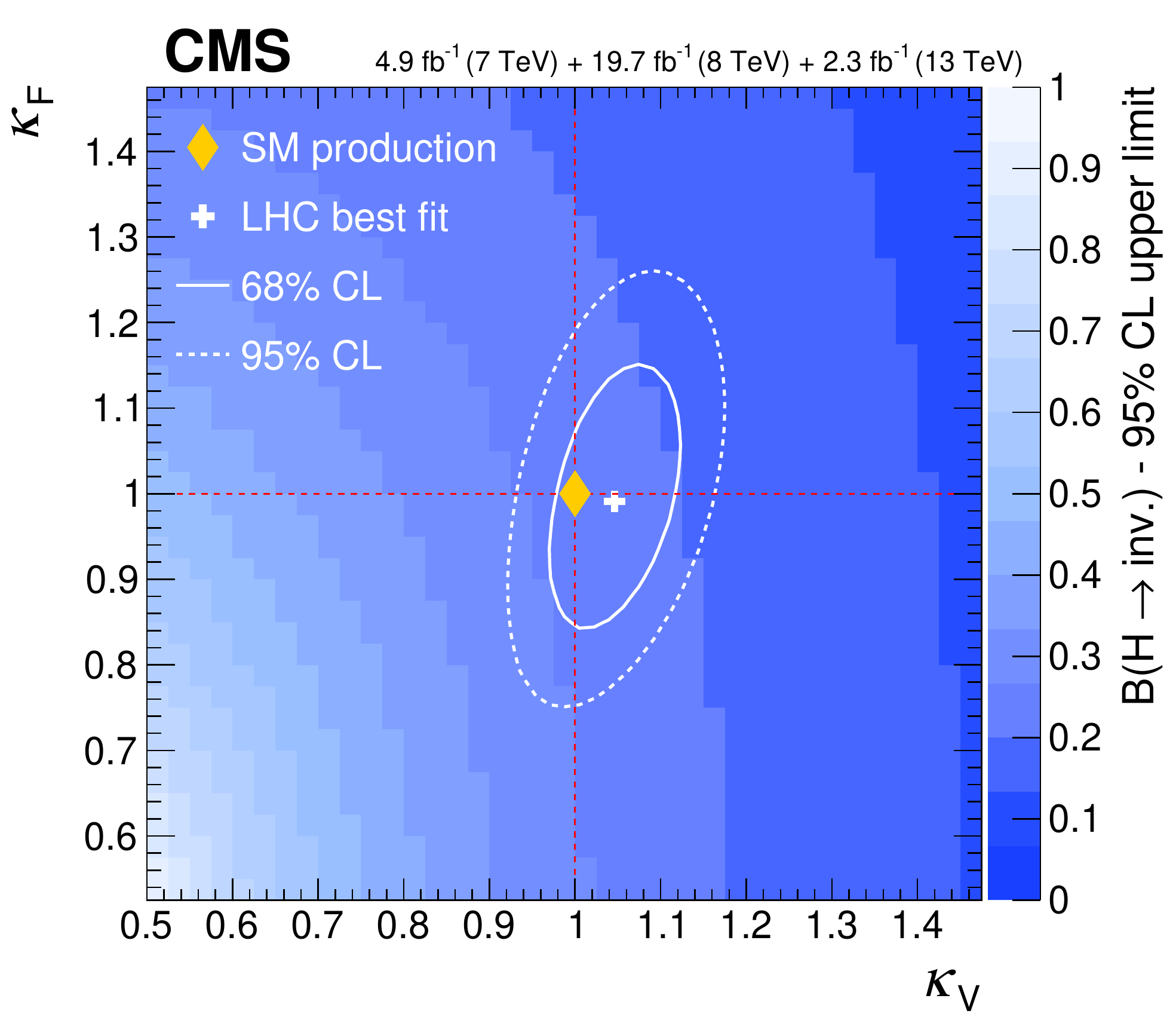}
    \caption{Observed 95\% CL upper limits on \brinv assuming a
    Higgs boson with a mass of 125\GeV whose production cross sections are scaled,
    relative to their SM values as a function of the coupling modifiers $\kappa_{F}$ and $\kappa_{V}$.
    The best-fit, and 68 and 95\% confidence level regions for $\kappa_{F}$ and $\kappa_{V}$ from
    Ref.~\cite{CMS-PAS-HIG-15-002} are superimposed as the solid and dashed white contours, respectively.
    The SM prediction (yellow diamond) corresponds to $\kappa_{F}=\kappa_{V}=1$.
    }
    \label{fig:2dlims}
\end{figure}

By varying the assumed SM production rates, the relative sensitivity of the different categories to an invisible
Higgs boson decay signal is studied. The rates for ggH, qqH, and VH production can be expressed in terms of the relative
coupling modifiers $\kappa_{F}$ and $\kappa_{V}$
that scale the couplings of the Higgs boson to the SM fermions and vector bosons, respectively~\cite{Heinemeyer:2013tqa}.
In this formalism, the total width of the Higgs boson is the sum of the partial widths to the visible channels, determined as
a function of $\kappa_{V}$ and $\kappa_{F}$, and an invisible decay width. The contribution from the $\Pg\Pg\to\Z\PH$ mode is scaled to account for the
interference between the $\PQt\PH$ and $\Z\PH$ diagrams (see Fig.~\ref{fig:production_ggZH}). The background from $\zvv\PH(\bbbar)$ production in the $\zbb$ search
is scaled consistently
with the other search channels. The SM production rates are recovered for $\kappa_{F}=\kappa_{V}=1$.
Figure~\ref{fig:2dlims} shows 95\% CL upper limits on \brinv obtained as a function of $\kappa_{F}$ and $\kappa_{V}$.
The best-fit, and 68 and 95\% CL limits for $\kappa_{F},~\kappa_{V}$ from Ref.~\cite{CMS-PAS-HIG-15-002} are superimposed. The observed upper
limit on \brinv varies between 0.18 and 0.29 within the 95\% confidence level region shown. An alternative model under which the
production rates are varied is described in Appendix~\ref{app:appendix}.

\begin{figure}[hbt]
  \centering
  \includegraphics[width=1.5\cmsFigWidth]{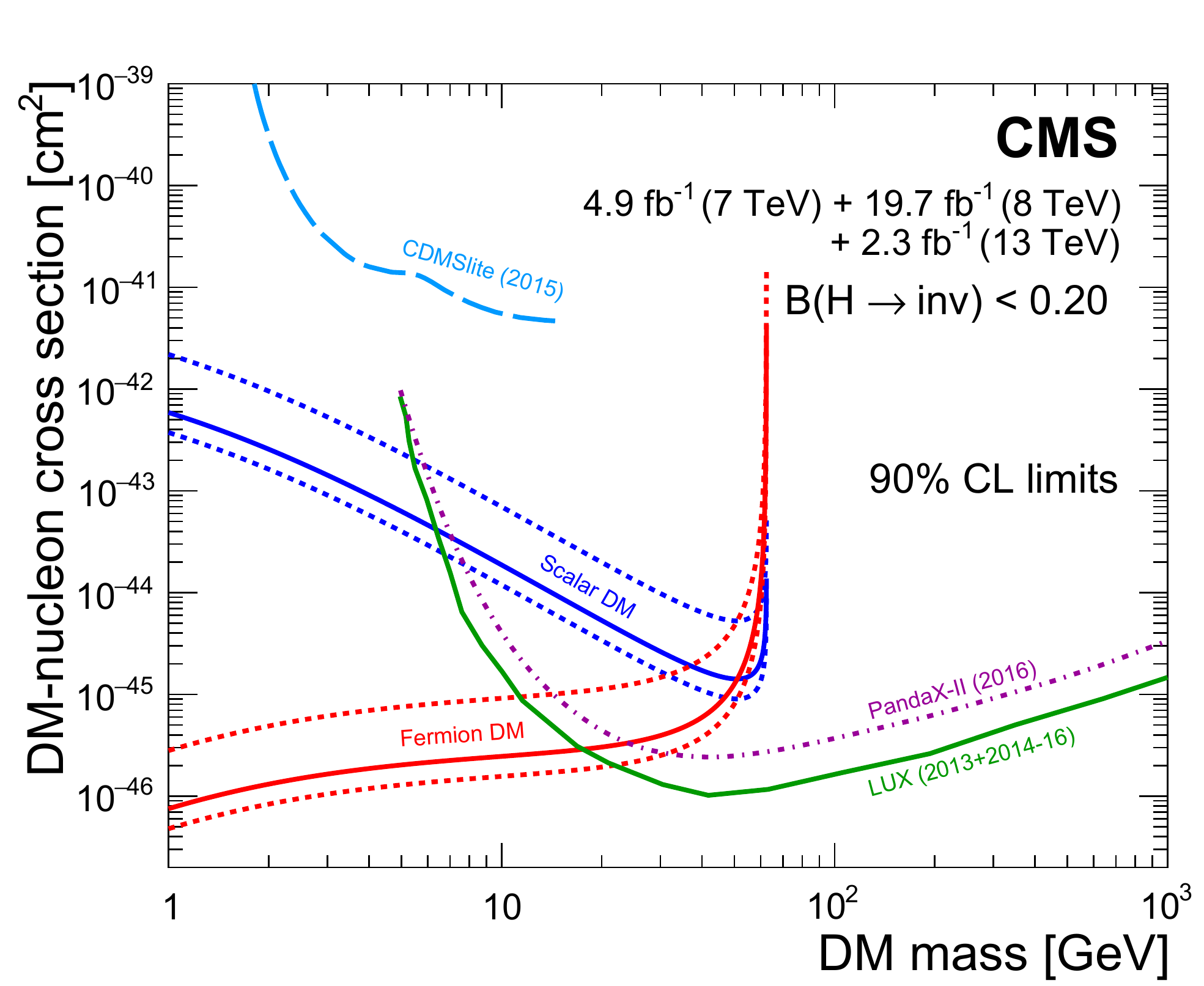}
\caption{
    Limits on the spin-independent DM-nucleon scattering cross section in Higgs-portal models assuming a
    scalar or fermion DM particle. The dashed lines show the variation in the exclusion limit using alternative values for $f_{\mathrm{N}}$ as
    described in the text.
    The limits are given at the 90\% CL to allow for comparison
    to direct detection constraints from the LUX~\cite{Akerib:2016vxi},
    PandaX-II~\cite{Tan:2016zwf}, and CDMSlite~\cite{PhysRevLett.116.071301} experiments.
    }
  \label{fig:hportal}
\end{figure}

The upper limit on \brinv, under the assumption of SM production cross sections for the Higgs boson, can be interpreted
in the context of a Higgs-portal model of DM interactions. In these models, a hidden sector provides a stable DM particle
candidate with tree-level couplings to the SM Higgs sector. Direct detection experiments are sensitive to elastic interactions
between DM particles and nuclei via Higgs boson exchange. These interactions produce nuclear recoil signatures, which can
be interpreted in terms of a DM-nucleon interaction cross section. The sensitivity varies as a function of the DM particle mass $m_{\chi}$ with
relatively small DM masses being harder to probe.
If the DM mass is smaller than $m_{\PH}/2$, the invisible Higgs boson decay width, $\Ginv$, can be translated via an effective field
theory approach into the spin-independent
DM-nucleon elastic cross section $\sigma^{\mathrm{SI}}$, assuming either a scalar or fermion DM candidate~\cite{Djouadi:2011aa}.
The translation is given by
\begin{equation}
\sigma^\mathrm{SI}_{\mathrm{S}-\mathrm{N}} = \frac{4\Ginv}{m_{\PH}^3v^2\beta} \frac{m_\mathrm{N}^4f_\mathrm{N}^2}{(m_\chi+m_\mathrm{N})^2},
\end{equation}
assuming a scalar DM candidate, and
\begin{equation}
\sigma^\mathrm{SI}_{\mathrm{f}-\mathrm{N}} = \frac{8\Ginv m_\chi^2}{m_{\PH}^5v^2\beta^3}\frac{m_\mathrm{N}^4f_\mathrm{N}^2}{(m_\chi+m_\mathrm{N})^2},
\end{equation}
assuming a fermion DM candidate, where $m_{\mathrm{N}}$ is the average of the proton and neutron masses 0.939\GeV and $\beta=\sqrt{1-4m^{2}_{\chi}/{m_{\PH}}^2}$.
The Higgs vacuum expectation value $v$ is taken to be 246\GeV.
The dimensionless quantity $f_{\mathrm{N}}$ denotes the nuclear form-factor.
The central values for the exclusion limits are derived assuming
$f_{\mathrm{N}}=0.326$, taken from Ref.~\cite{PhysRevD.81.014503}, while alternative values of 0.260 and 0.629 are taken from the
MILC Collaboration~\cite{PhysRevLett.103.122002}.
The translation between $\Ginv$ and $\brinv$ uses the relation $\brinv=\Ginv/(\Gsm+\Ginv)$, where $\Gsm=4.07$\MeV~\cite{Heinemeyer:2013tqa}.
Figure~\ref{fig:hportal} shows the 90\% CL upper limits on the spin-independent DM-nucleon cross section as a function of the
DM mass, assuming $m_{\PH}=125$\GeV, for the scalar and fermion DM scenarios. These limits are calculated using the 90\% CL
limit of $\brinv<\higgsbrobsninety$ in
order to compare with those from  the LUX~\cite{Akerib:2016vxi}, PandaX-II~\cite{Tan:2016zwf}, and
CDMSlite~\cite{PhysRevLett.116.071301} experiments,
which provide the strongest direct constraints on the spin-independent DM-nucleon cross section in the range of DM particle
masses probed by this analysis.
Under the assumptions of the Higgs-portal models, the present CMS results
provide more stringent limits for DM masses below roughly 20 or 5\GeV, assuming a fermion or scalar DM particle, respectively.

\section{Summary}

A combination of searches for a Higgs boson decaying to invisible particles using proton-proton collision data collected during 2011, 2012, and 2015,
at centre-of-mass energies of 7, 8, and 13\TeV, respectively, is presented.
The combination includes searches targeting Higgs boson production in the ZH mode, in which a Z boson decays to
$\ell^{+}\ell^{-}$ or \bbbar, and the qqH mode, which is the most sensitive channel. The combination also
includes the first searches at CMS targeting VH production, in which the vector boson decays hadronically,
and the ggH mode in which the Higgs boson is produced in association with jets. No significant deviations from the SM predictions
are observed and upper limits are placed on the branching fraction for the Higgs boson decay to invisible particles. The combination of all searches
yields an observed (expected) upper limit on \brinv of $\higgsbrobs$ $(\higgsbrexp)$ at the 95\% confidence level, assuming SM production of the Higgs boson.
The combined 90\% confidence level limit of $\brinv<\higgsbrobsninety$ has been interpreted in Higgs-portal models and constraints are placed on the spin-independent DM-nucleon interaction cross section.
These limits provide stronger constraints than those from direct detection experiments for DM masses
below roughly 20\,(5)\GeV, assuming a fermion (scalar) DM particle, within the context of Higgs-portal models.

\clearpage
\begin{acknowledgments}
\hyphenation{Bundes-ministerium Forschungs-gemeinschaft Forschungs-zentren Rachada-pisek} We congratulate our colleagues in the CERN accelerator departments for the excellent performance of the LHC and thank the technical and administrative staffs at CERN and at other CMS institutes for their contributions to the success of the CMS effort. In addition, we gratefully acknowledge the computing centres and personnel of the Worldwide LHC Computing Grid for delivering so effectively the computing infrastructure essential to our analyses. Finally, we acknowledge the enduring support for the construction and operation of the LHC and the CMS detector provided by the following funding agencies: the Austrian Federal Ministry of Science, Research and Economy and the Austrian Science Fund; the Belgian Fonds de la Recherche Scientifique, and Fonds voor Wetenschappelijk Onderzoek; the Brazilian Funding Agencies (CNPq, CAPES, FAPERJ, and FAPESP); the Bulgarian Ministry of Education and Science; CERN; the Chinese Academy of Sciences, Ministry of Science and Technology, and National Natural Science Foundation of China; the Colombian Funding Agency (COLCIENCIAS); the Croatian Ministry of Science, Education and Sport, and the Croatian Science Foundation; the Research Promotion Foundation, Cyprus; the Secretariat for Higher Education, Science, Technology and Innovation, Ecuador; the Ministry of Education and Research, Estonian Research Council via IUT23-4 and IUT23-6 and European Regional Development Fund, Estonia; the Academy of Finland, Finnish Ministry of Education and Culture, and Helsinki Institute of Physics; the Institut National de Physique Nucl\'eaire et de Physique des Particules~/~CNRS, and Commissariat \`a l'\'Energie Atomique et aux \'Energies Alternatives~/~CEA, France; the Bundesministerium f\"ur Bildung und Forschung, Deutsche Forschungsgemeinschaft, and Helmholtz-Gemeinschaft Deutscher Forschungszentren, Germany; the General Secretariat for Research and Technology, Greece; the National Scientific Research Foundation, and National Innovation Office, Hungary; the Department of Atomic Energy and the Department of Science and Technology, India; the Institute for Studies in Theoretical Physics and Mathematics, Iran; the Science Foundation, Ireland; the Istituto Nazionale di Fisica Nucleare, Italy; the Ministry of Science, ICT and Future Planning, and National Research Foundation (NRF), Republic of Korea; the Lithuanian Academy of Sciences; the Ministry of Education, and University of Malaya (Malaysia); the Mexican Funding Agencies (BUAP, CINVESTAV, CONACYT, LNS, SEP, and UASLP-FAI); the Ministry of Business, Innovation and Employment, New Zealand; the Pakistan Atomic Energy Commission; the Ministry of Science and Higher Education and the National Science Centre, Poland; the Funda\c{c}\~ao para a Ci\^encia e a Tecnologia, Portugal; JINR, Dubna; the Ministry of Education and Science of the Russian Federation, the Federal Agency of Atomic Energy of the Russian Federation, Russian Academy of Sciences, and the Russian Foundation for Basic Research; the Ministry of Education, Science and Technological Development of Serbia; the Secretar\'{\i}a de Estado de Investigaci\'on, Desarrollo e Innovaci\'on and Programa Consolider-Ingenio 2010, Spain; the Swiss Funding Agencies (ETH Board, ETH Zurich, PSI, SNF, UniZH, Canton Zurich, and SER); the Ministry of Science and Technology, Taipei; the Thailand Center of Excellence in Physics, the Institute for the Promotion of Teaching Science and Technology of Thailand, Special Task Force for Activating Research and the National Science and Technology Development Agency of Thailand; the Scientific and Technical Research Council of Turkey, and Turkish Atomic Energy Authority; the National Academy of Sciences of Ukraine, and State Fund for Fundamental Researches, Ukraine; the Science and Technology Facilities Council, UK; the US Department of Energy, and the US National Science Foundation.

Individuals have received support from the Marie-Curie programme and the European Research Council and EPLANET (European Union); the Leventis Foundation; the A. P. Sloan Foundation; the Alexander von Humboldt Foundation; the Belgian Federal Science Policy Office; the Fonds pour la Formation \`a la Recherche dans l'Industrie et dans l'Agriculture (FRIA-Belgium); the Agentschap voor Innovatie door Wetenschap en Technologie (IWT-Belgium); the Ministry of Education, Youth and Sports (MEYS) of the Czech Republic; the Council of Science and Industrial Research, India; the HOMING PLUS programme of the Foundation for Polish Science, cofinanced from European Union, Regional Development Fund, the Mobility Plus programme of the Ministry of Science and Higher Education, the National Science Center (Poland), contracts Harmonia 2014/14/M/ST2/00428, Opus 2013/11/B/ST2/04202, 2014/13/B/ST2/02543 and 2014/15/B/ST2/03998, Sonata-bis 2012/07/E/ST2/01406; the Thalis and Aristeia programmes cofinanced by EU-ESF and the Greek NSRF; the National Priorities Research Program by Qatar National Research Fund; the Programa Clar\'in-COFUND del Principado de Asturias; the Rachadapisek Sompot Fund for Postdoctoral Fellowship, Chulalongkorn University and the Chulalongkorn Academic into Its 2nd Century Project Advancement Project (Thailand); and the Welch Foundation, contract C-1845.
\end{acknowledgments}

\bibliography{auto_generated}

\appendix
\section{Supplementary material}

\label{app:appendix}

\subsection{Negative likelihood scans}

\begin{figure}[hbt]
  \centering
  \includegraphics[width=1.2\cmsFigWidth]{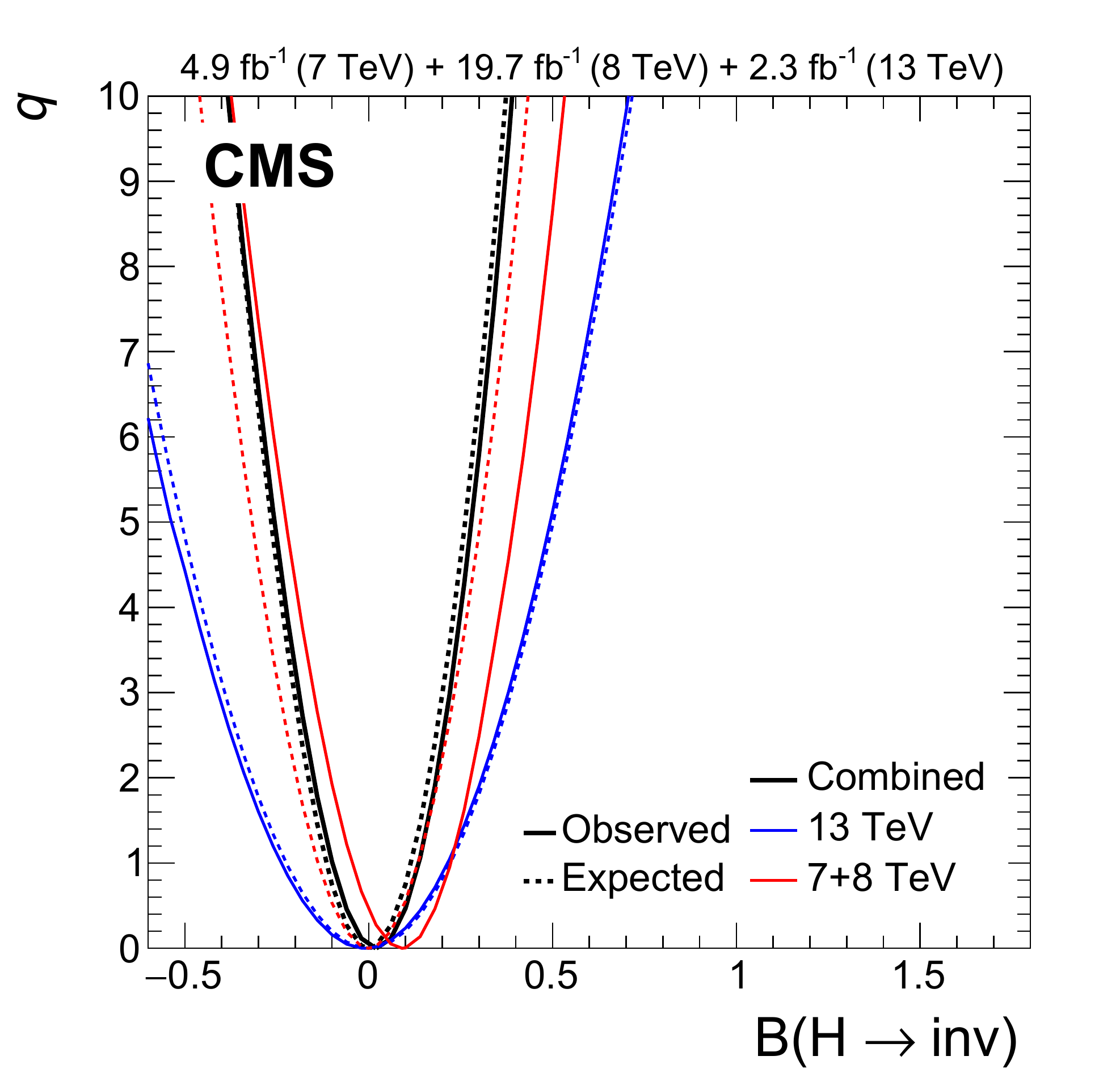}
  \includegraphics[width=1.2\cmsFigWidth]{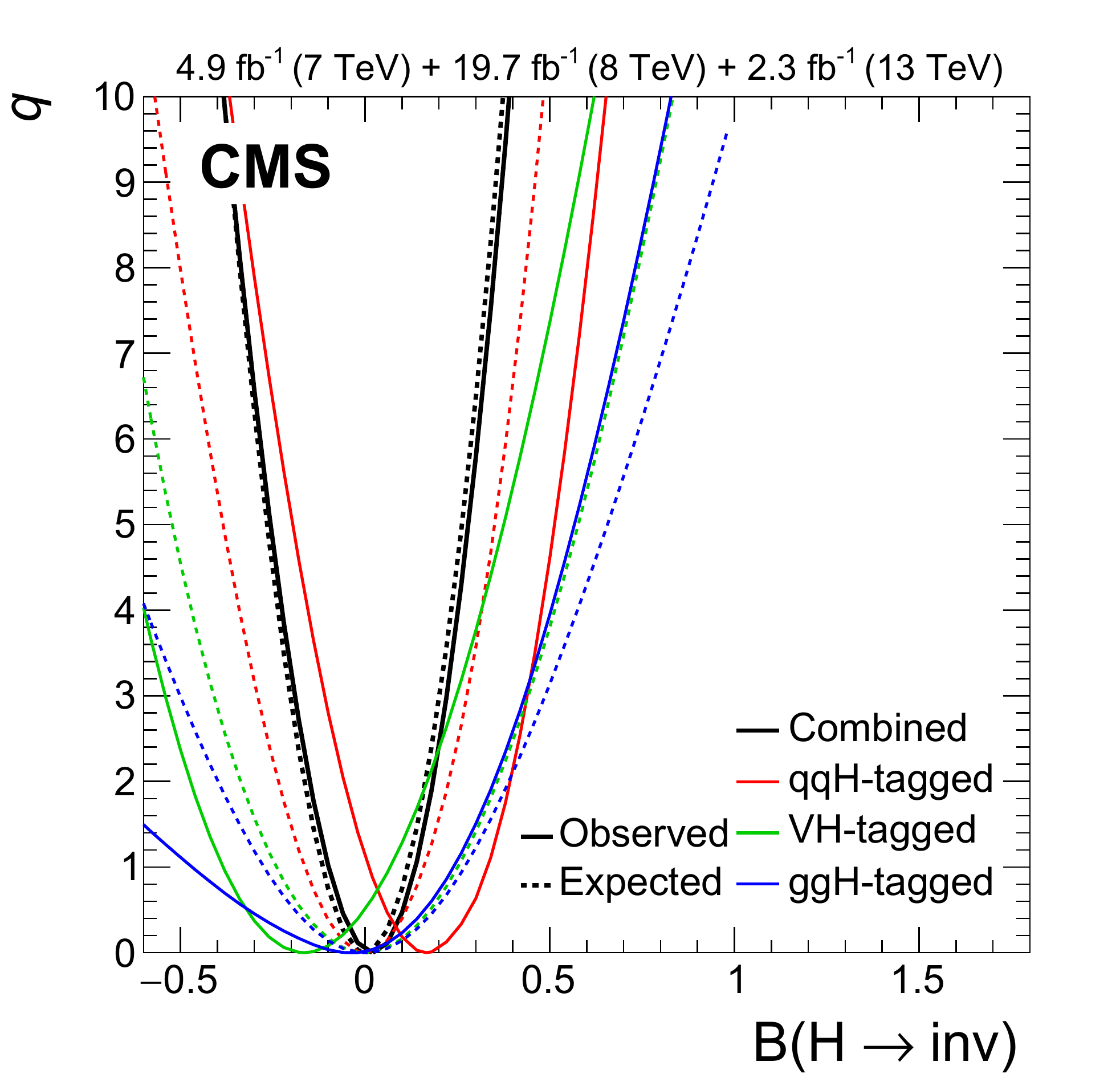}
\caption{Profile likelihood ratio as a function of \brinv assuming SM production
    cross sections of a Higgs boson with a mass of 125\GeV.
    The solid curves represent the observations in data and the
    dashed curves represent the expected result assuming no
    invisible decays of the Higgs boson.
    (left) The observed and expected likelihood scans for the partial combinations
    of the 7+8 and 13\TeV analyses, and the full combination.
    (right) The observed and expected likelihood scans for the partial combinations
    of the VBF-tagged, VH-tagged, and ggH-tagged analyses, and the full combination.
    }
  \label{fig:lhscan_neg}
\end{figure}

The profile likelihood ratio as a function of \brinv using partial combinations of the 7+8 and 13\TeV analyses, and for the
full combination are shown in Fig.~\ref{fig:lhscan_neg}\,(left).
The profile likelihood ratio scans for the partial combinations of the VBF-tagged,
VH-tagged, and ggH-tagged analyses are shown in Fig.~\ref{fig:lhscan_neg}\,(right).
The results are shown for the data and for an Asimov
data set~\cite{Cowan:2010js} in which $\brinv=0$ is assumed. For these results, the
condition that $\brinv>0$ is removed.

\subsection{Non-SM production cross sections}

\begin{figure}[hbt]
\centering
  \includegraphics[width=1.2\cmsFigWidth]{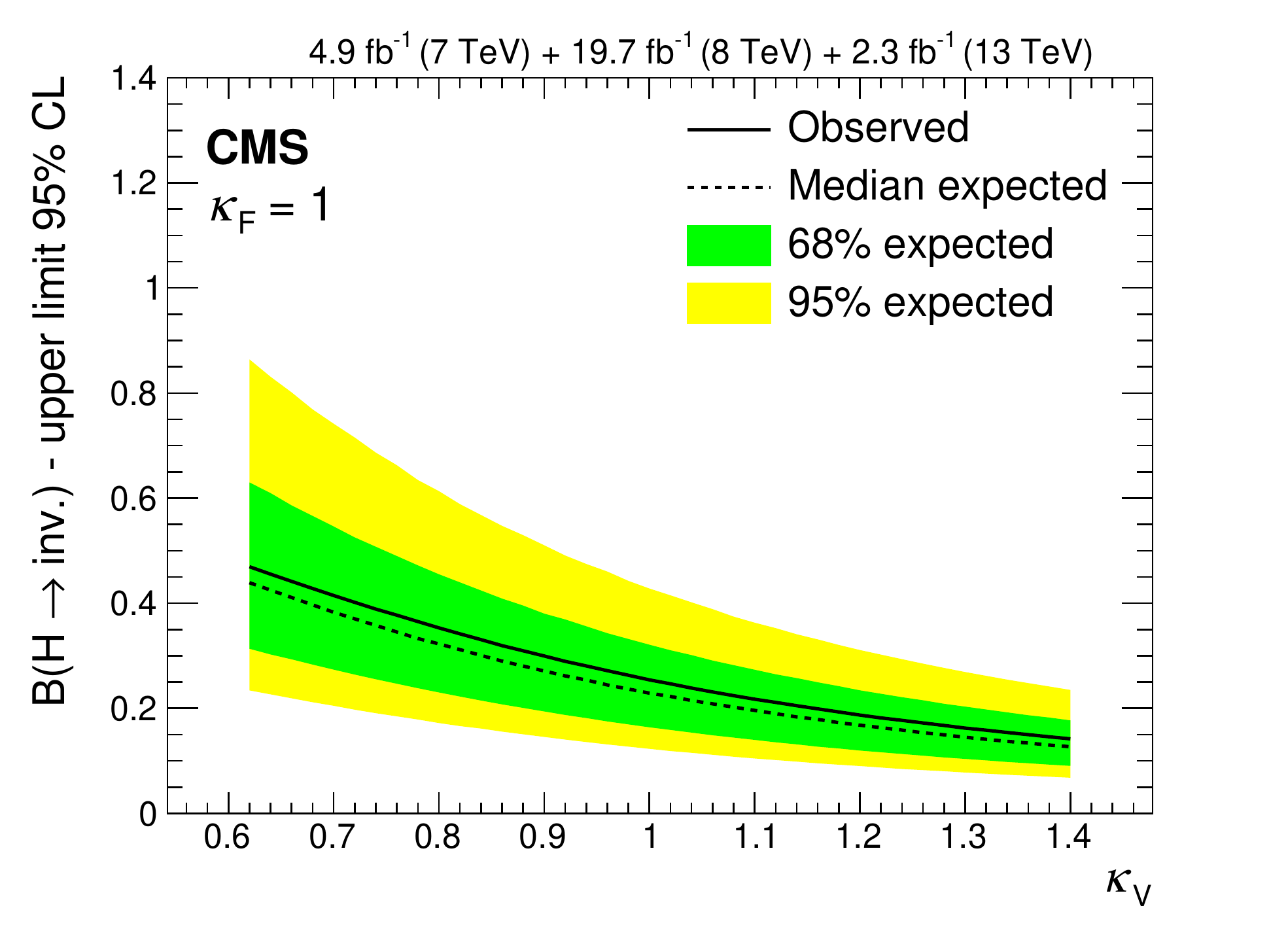}
  \includegraphics[width=1.2\cmsFigWidth]{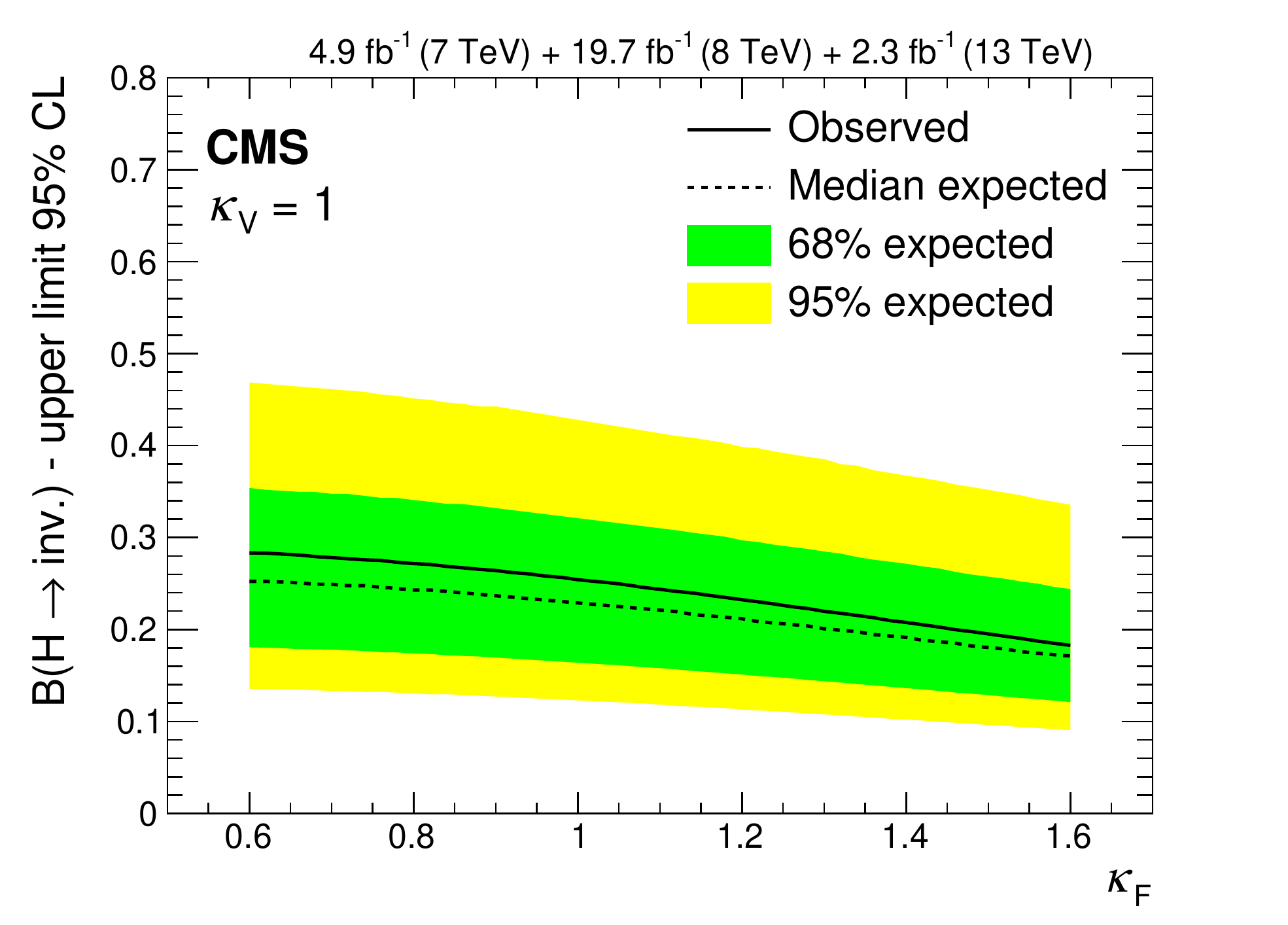}
    \caption{Observed and expected 95\% CL upper limits on \brinv assuming a
    Higgs boson with a mass of 125\GeV whose production cross sections are scaled,
    relative to their SM values as a function of (left) $\kappa_{V}$, fixing $\kappa_{F}=1$ and (right)
    $\kappa_{F}$, fixing $\kappa_{V}=1$.
    }
    \label{fig:1dlims_kvkf}
\end{figure}

Figure~\ref{fig:1dlims_kvkf} shows the observed and expected 95\% CL upper limits on \brinv obtained as a function of
either $\kappa_{V}$, fixing $\kappa_{F}=1$ or as a function of $\kappa_{F}$, fixing $\kappa_{V}=1$.

\begin{figure}[hbt]
\centering
  \includegraphics[width=1.5\cmsFigWidth]{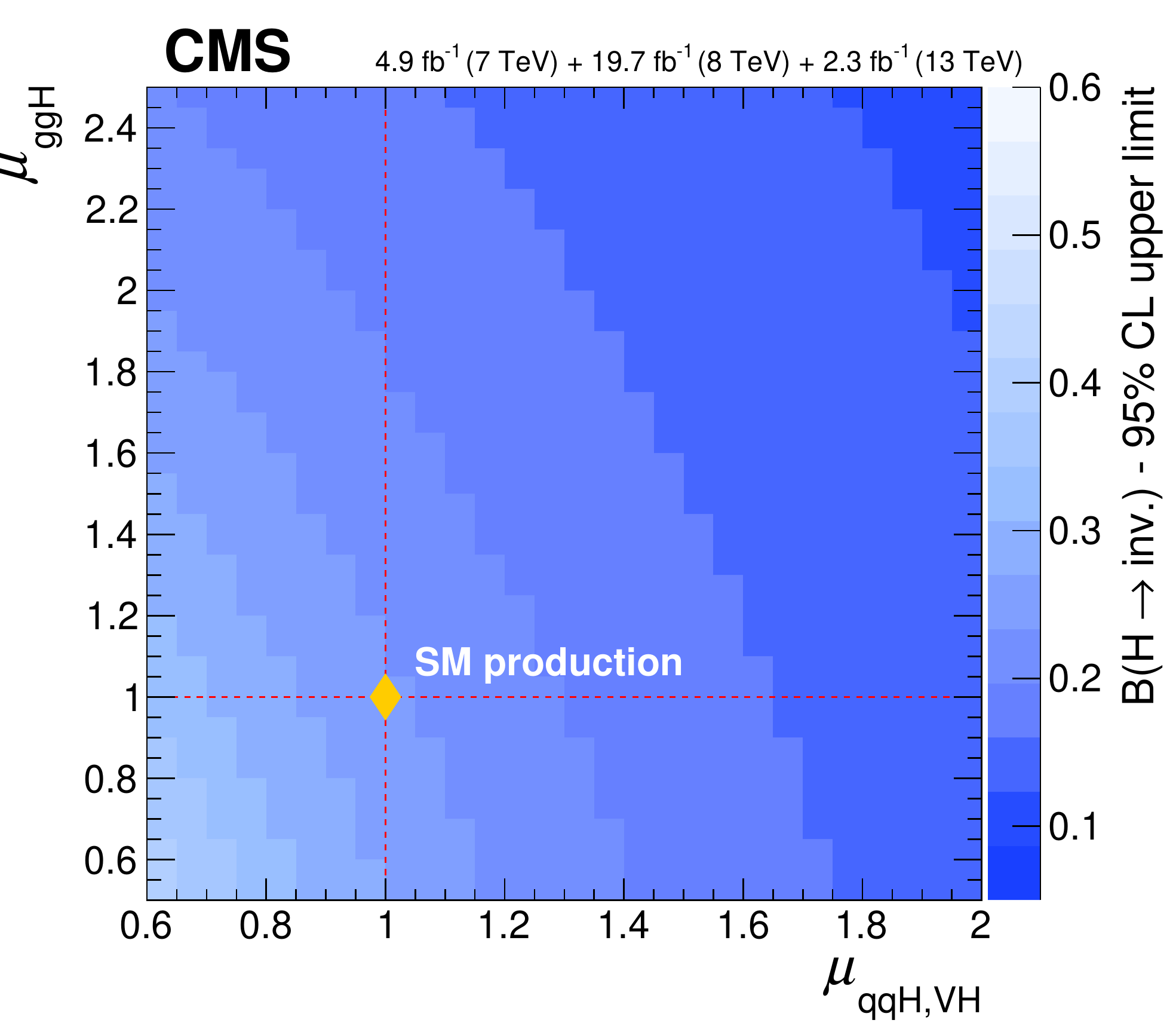}
    \caption{Observed 95\% CL upper limits on \brinv assuming a
    Higgs boson with a mass of 125\GeV whose production cross sections are scaled,
    relative to their SM values, by  $\mu_{\Pg\Pg\PH}$ and $\mu_{\PQq\PQq\PH,\mathrm{V}\PH}$.
    The SM (yellow diamond) is attained for $\mu_{\Pg\Pg\PH}=\mu_{\PQq\PQq\PH,\mathrm{V}\PH}=1$.}
    \label{fig:2dlims_mu}
\end{figure}

The rates for the different production modes can be scaled by the multiplicative factors $\mu_{\Pg\Pg\PH}$ and $\mu_{\PQq\PQq\PH,\mathrm{V}\PH}$
which respectively denote the production cross section values for the ggH and qqH/VH modes
relative to their SM predictions. The SM production cross sections are
therefore attained for $\mu_{\Pg\Pg\PH}=\mu_{\PQq\PQq\PH,\mathrm{V}\PH}=1$.
Figure~\ref{fig:2dlims_mu} shows the 95\% CL upper limits on \brinv obtained as a function of  $\mu_{\Pg\Pg\PH}$ and $\mu_{\PQq\PQq\PH,\mathrm{V}\PH}$.

\subsection{Uncertainty breakdown}

\begin{figure}[hbt]
  \centering
 \includegraphics[width=1.5\cmsFigWidth]{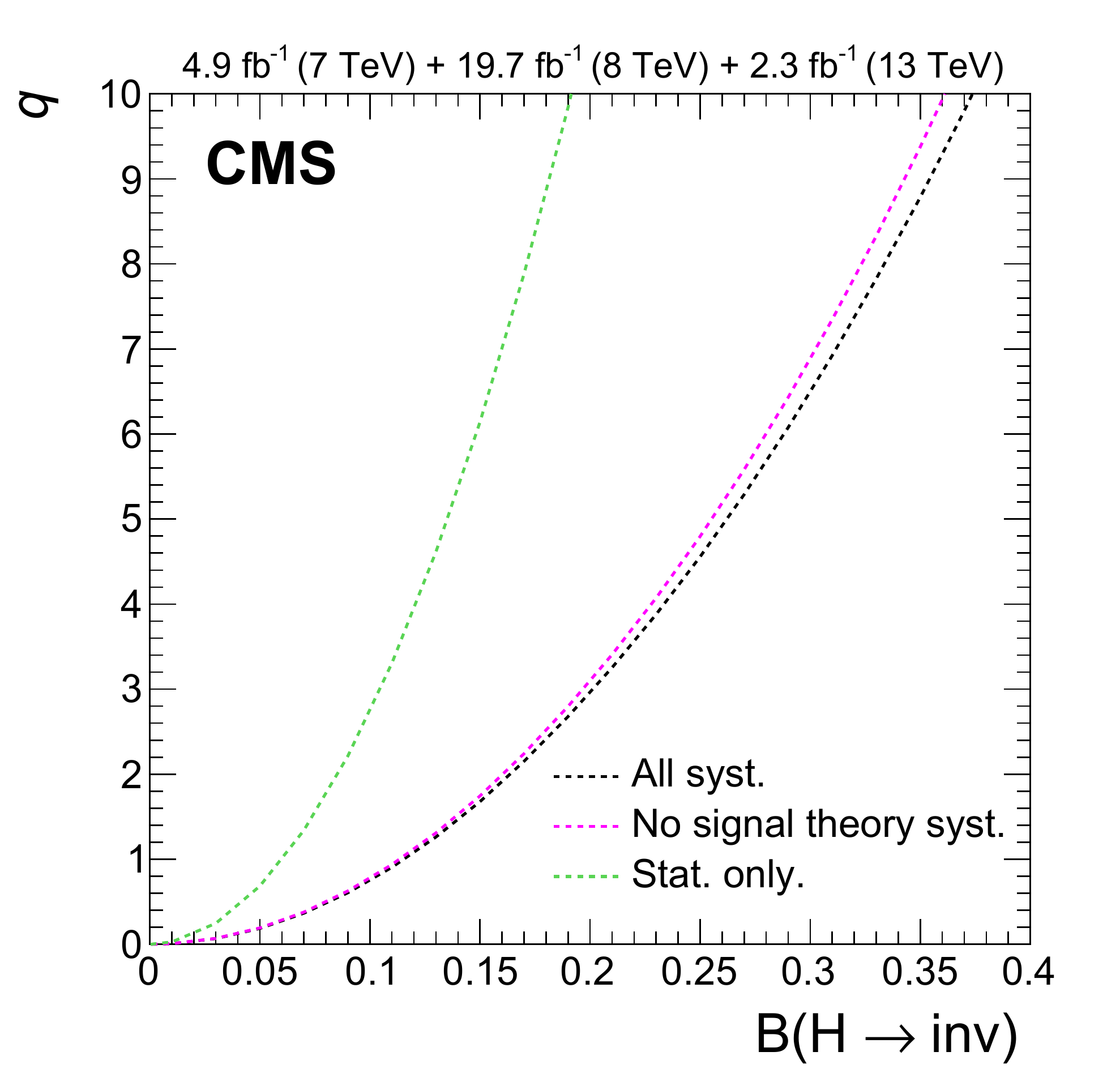}
\caption{
    Expected profile likelihood ratio as a function of \brinv assuming SM production
    cross sections of a Higgs boson with a mass of 125\GeV. The results fixing all nuisance
    parameters associated to theoretical systematic uncertainties on the signal to their
    nominal values in data is shown as the magenta line. The result assuming only
    statistical uncertainties is also shown in green.
    }
  \label{fig:lhscan_app}
\end{figure}

The profile likelihood ratio using the Asimov dataset fixing all nuisance parameters associated
with theoretical systematic uncertainties in the signal model to their nominal
values from the combined fit to data is shown in Fig.~\ref{fig:lhscan_app}.
A second result including only statistical uncertainties is additionally shown.

\cleardoublepage \section{The CMS Collaboration \label{app:collab}}\begin{sloppypar}\hyphenpenalty=5000\widowpenalty=500\clubpenalty=5000\input{HIG-16-016-authorlist.tex}\end{sloppypar}
\end{document}

%% file: HIG-16-016-authorlist.tex
\textbf{Yerevan Physics Institute,  Yerevan,  Armenia}\\*[0pt]
V.~Khachatryan, A.M.~Sirunyan, A.~Tumasyan
\vskip\cmsinstskip
\textbf{Institut f\"{u}r Hochenergiephysik,  Wien,  Austria}\\*[0pt]
W.~Adam, E.~Asilar, T.~Bergauer, J.~Brandstetter, E.~Brondolin, M.~Dragicevic, J.~Er\"{o}, M.~Flechl, M.~Friedl, R.~Fr\"{u}hwirth\cmsAuthorMark{1}, V.M.~Ghete, C.~Hartl, N.~H\"{o}rmann, J.~Hrubec, M.~Jeitler\cmsAuthorMark{1}, A.~K\"{o}nig, I.~Kr\"{a}tschmer, D.~Liko, T.~Matsushita, I.~Mikulec, D.~Rabady, N.~Rad, B.~Rahbaran, H.~Rohringer, J.~Schieck\cmsAuthorMark{1}, J.~Strauss, W.~Waltenberger, C.-E.~Wulz\cmsAuthorMark{1}
% ...
\vskip\cmsinstskip
% ...
1:~~Also at Vienna University of Technology, Vienna, Austria\\
% ...
